\pdfoutput=1
\documentclass[preprint,12pt,authoryear]{elsarticle}

\usepackage{amssymb}
\usepackage{lineno}
\usepackage{float}
\usepackage{graphicx}
\usepackage{rotating}
\usepackage{cases}
\journal{Icarus}

\begin{document}

\begin{frontmatter}



\title{Eight-year Climatology of Dust Optical Depth on Mars}


\author[label1,label2,label3]{L.~Montabone\corref{cor1}}
\ead{lmontabone@SpaceScience.org}
\cortext[cor1]{Laboratoire de M\'et\'eorologie Dynamique, Universit\'e Pierre et
Marie Curie, Tour 45-55, 3\`eme \'etage, 4, place Jussieu, 75252 Paris Cedex 05,
France.}
\author[label1]{F.~Forget}
\author[label1]{E.~Millour}
\author[label4]{R.~J.~Wilson}
\author[label5]{S.~R.~Lewis}
\author[label6]{B.~A.~Cantor}
\author[label7]{D.~Kass}
\author[label7]{A.~Kleinb\"ohl}
\author[label8]{M.~T.~Lemmon}
\author[label9]{M.~D.~Smith}
\author[label10]{M.~J.~Wolff}
\address[label1]{Laboratoire de M\'et\'eorologie Dynamique, Universit\'e Pierre
et Marie Curie, Paris, France.}
\address[label2]{Department of Physics, University of Oxford, Oxford, UK.}
\address[label3]{Space Science Institute, Boulder, CO, USA.}
\address[label4]{GFDL, Princeton, NJ, USA.}
\address[label5]{Department of Physical Sciences, The Open University, UK.}
\address[label6]{Malin Space Science Systems, San Diego, CA, USA.}
\address[label7]{JPL, Pasadena, CA, USA.}
\address[label8]{Texas A\&M University, College Station, TX, USA.}
\address[label9]{NASA Goddard Space Flight Center, Greenbelt, MD, USA.}
\address[label10]{Space Science Institute, Boulder, CO, USA.}

\begin{abstract}
We have produced a multiannual climatology of airborne dust from Martian year
24 to 31 using multiple datasets of retrieved or estimated column optical
depths. The datasets are based on observations of the Martian atmosphere from
April 1999 to July 2013 made by different orbiting instruments: the Thermal Emission
Spectrometer (TES) aboard Mars Global Surveyor, the Thermal Emission
Imaging System (THEMIS) aboard Mars Odyssey, and the Mars Climate Sounder (MCS)
aboard Mars Reconnaissance Orbiter (MRO). The procedure we have adopted
consists of gridding the available retrievals of column dust optical depth
(CDOD) from TES and THEMIS nadir observations, as well as the estimates of
this quantity from MCS limb observations. Our gridding method
calculates averages on a regularly spaced, but possibly incomplete, spatio-temporal grid, using an iterative procedure weighted in space, time, and retrieval uncertainty. In order to
evaluate strengths and weaknesses of the resulting gridded maps, we associate
values of 
weighted standard deviation with every grid point average, and compare with
independent observations of CDOD by PanCam cameras and Mini-TES spectrometers aboard the Mars Exploration
Rovers (``Spirit'' and ``Opportunity''), as well as the Compact Reconnaissance
Imaging Spectrometer for Mars aboard MRO. We have statistically
analyzed the irregularly gridded maps to provide an overview of the dust climatology on Mars
over eight years, specifically in relation to its interseasonal and interannual
variability. Finally, we have produced multiannual, regular daily maps of CDOD
by spatially interpolating the irregularly gridded maps using a kriging method. These synoptic maps are used as dust scenarios in the
Mars Climate Database version 5, and are useful in many modelling
applications in addition to forming a basis for instrument intercomparisons. 
The derived dust maps for the eight available Martian years (currently version 1.5) are publicly available and distributed with open access.

\end{abstract}


\end{frontmatter}


\section{Introduction}
%%%%%%%%%%%%%%%%%%%%%%%%%%
\label{intro}

The dust cycle is currently considered as the key process controlling the
variability of the Martian climate at interseasonal and interannual time scales,
as well as the weather variability at much shorter time scales. The atmospheric
thermal and dynamical structures, as well as the transport of aerosols and
chemical species, are all strongly dependent on the dust spatio-temporal
distribution, particle sizes, and optical properties.

Since the first scientific observations of a planet-encircling dust storm by
ground-based telescopes in the late 50's, dust has been one of the main
objectives of many of the spacecraft missions to Mars over more than 40 years. Recent and ongoing missions, 
such as Mars Global Surveyor (MGS), Mars Odyssey (ODY), Mars Express, Mars
Exploration Rover (MER), Mars Reconnaissance Orbiter (MRO), and Mars Science Laboratory,  
%\citep[MGS, ][[]{}, Mars Odyssey \citep[ODY, ][]{}, Mars Express, Mars Exploration Rover \citep[MER, ][]{}, Mars Reconnaissance Orbiter \citep[MRO, ][]{}, and Mars Science Laboratory, 
have included spectrometers, radiometers or imagers to measure radiances at wavelengths sensitive to dust.  

Atmospheric dust can be qualitatively observed by spacecraft
cameras, such as the Mars Orbiter Camera (MOC) aboard MGS, and the Mars Color Imager (MARCI)
aboard MRO \citep[See][for dust climatologies based on camera
observations, as well as the Mars Daily Weather websites\footnote{\texttt{http://www.msss.com/mars\_images/moc/weather\_reports/} for MOC,\\ \texttt{http://www.msss.com/msss\_images/latest\_weather.html} for MARCI.}]{can01, can07, can08, wan13}. One of the key physical parameters used to quantify the
presence of mineral dust in the atmosphere is the column (or total) optical depth (or
optical thickness). It is a measure of the
fraction of radiation at a specific wavelength that would be removed from the vertical component of a beam
during its path through the atmosphere by
absorption and scattering due to airborne dust. It can also be defined as the
integral over an atmospheric column of the profile of the extinction opacity
(or extinction coefficient), where extinction accounts for absorption and
scattering.

The column optical depth is the product of dust retrievals when the
radiances are obtained by nadir-viewing instruments. Profiles of extinction
opacity can be derived from radiances measured by limb-viewing instruments,
providing important information on the vertical extension of the dust. Other
important properties related to dust, such as the size distribution and the
optical parameters, are more difficult to retrieve from direct measurements of
radiances \citep{cla03,wol03,wol06,wol09}. The knowledge of the spatio-temporal
distribution of dust is of primary importance to produce quantitative estimates
of dust mass mixing ratios, and calculate the atmospheric heating rates due to
absorption and scattering of solar and infrared (IR) radiation by airborne particles.
These calculations are the basis for describing the thermal forcing in Mars
atmospheric models, and producing accurate predictions of the atmospheric state.

The choice of scenarios (i.e. spatio-temporal distributions) of dust optical depth has a significant impact on Martian global climate model (GCM) simulations. Model studies have often been carried out with analytical specifications of dust distributions, both in the horizontal and in the vertical \citep{for99,mont04,kur08}. More recently, modeling groups have been carrying out simulations with more realistic horizontal dust distributions, tied to TES observations \citep{guz13a,wan13,kav13}. We propose in this paper to create a well-documented set of dust scenarios to be used in model experiments. Although we focus our attention on the column-integrated dust distribution, it should be noted that the spatial variation of the vertical distribution of dust also plays a significant role in thermal response \citep[see e.g.][]{guz13a}. There are a number of approaches to represent this vertical structure, which is beyond the scope of this paper.    

To date, there exist several datasets of retrieved column dust optical depth (CDOD) for Mars, spanning more
than 20 Martian years (MY) since the Mariner era. These datasets are
highly heterogeneous, as they have been created using data from different
instruments having different geometric views, spatial and temporal coverage, as
well as different observing wavelengths. Nonetheless, we show in this paper that
at least some of these datasets can be appropriately used to quantitatively
reconstruct the recent dust climatology on Mars, and characterize the variability over
many seasonal cycles. This paper seeks to produce a continuous, multiannual climatology of CDOD from early March 1999 (solar longitude
$\mathrm{L_{s}}\sim104^{\circ}$ in MY 24) to the end of July 2013 ($\mathrm{L_{s}}=360^{\circ}$ in MY 31).
During this period of time, the Thermal Emission Spectrometer \citep[TES, ][]{chr01} aboard MGS, the Thermal Emission
Imaging System \citep[THEMIS, ][]{chr04} on ODY, and the Mars Climate Sounder \citep[MCS, ][]{mcc07} on MRO provided global
coverage of radiance observations at IR wavelengths, from which \citet{smi03,smi04,smi09,kle09} obtained 
direct retrievals of CDOD or estimates of this quantity from the integrated
extinction profiles.  

While images from orbiting spacecraft can provide information over large areas
on the planet at any given time, observations of IR radiances from orbiting
instruments have a very discrete coverage in longitude and local time due to the
choice of orbit geometry. MGS and MRO, for instance, have sun-synchronous, nearly 2-hour polar orbits, which provide good latitude coverage but only sample about a dozen longitudes, usually at close to two fixed local times of day except when crossing the poles (dust can often only be retrieved in the daytime, restricting further to one local time).
Because dust storms on Mars have a wide range of
spatial and temporal scales \citep{can07, wan03, wan05}, these discrete observations
can affect the space-time representation of dust storm activity. Extrapolating
the data collected along orbit tracks to a broader range of local time and
longitude introduces even more biases. Synoptic maps produced using simple average binning may alter the representation of rapidly evolving dust distribution.

We have developed a gridding methodology that is specifically adapted to
heterogeneous observations, and to the discrete longitudinal/temporal coverage typical of
spacecraft data acquisition. The objective is to produce daily, regularly
gridded maps of absorption CDOD at 9.3~$\mu\mathrm{m}$ for several consecutive Martian years (dust opacity in absorption is less dependent on the particle size than in extinction). To achieve
the goal, we have adopted a two-step procedure. The first step consists in
calculating iteratively averages and standard deviations of observations on a regularly spaced but
likely incomplete spatio-temporal grid, after having binned the data using time windows of different size, and applied
appropriate weighting functions in space, time, and retrieval uncertainty at each iteration. We
have established acceptance criteria for the gridded values, based on the number of
reliable observations in each bin. We have used this first product to
statistically study the dust variability over almost eight complete Martian
years. The second step consists in producing regular, daily maps of CDOD
by spatially interpolating and/or extrapolating the irregularly gridded maps,
using a kriging method \citep[see e.g.][for a general introduction on the technique]{jou78}. This multiannual series of synoptic maps of CDOD is
used for the dust scenarios in the GCM simulations that
produce the current version of the Mars Climate Database \citep[MCD version 5.1,][]{mil14}.

We provide open access to both the irregularly gridded and regularly kriged datasets, to foster scientific analyses and applications of the long-term Martian dust climatology. The most up-to-date version of these products (currently v1.5) can be downloaded in the form of NetCDF files from the MCD project website\footnote{The URL to access the website
is: \texttt{http://www-mars.lmd.jussieu.fr/}}, hosted by
the Laboratoire de M\'et\'eorologie Dynamique (LMD). We also provide
atlases of these maps as supplementary material of
this publication.

The outline of this paper is the following. Section~\ref{instruments} describes
the instruments and data we have used to produce the dust climatology. We
provide details of the time coverage, data quality control,
processing, and uncertainties for each instrument. In Section~\ref{gridding} we introduce the iterative
weighted binning methodology we have adopted to create irregularly gridded (but regularly spaced) CDOD maps from the spacecraft observations.
Section~\ref{validation} is devoted to discuss the internal validation of the gridded
maps, as well as the validation with independent observations. We report in Section~\ref{analysis} the
statistical analysis of the dust climatology, in relation to the intraseasonal
and interannual variabilities. In Section~\ref{kriging} we discuss the
assumptions we have made and the kriging technique we have applied, with the purpose of
producing regular and complete maps to be used as dust scenarios. A summary is
outlined in Section~\ref{summary}, including a discussion on future developments.

%%%%%%%%%%%%%%%%%%%%%%%%%%
\section{Spacecraft, instruments and observations}
%%%%%%%%%%%%%%%%%%%%%%%%%%
\label{instruments}

Mars Global Surveyor started its science mapping phase in March 1999
($\mathrm{L_{s}}\sim104^{\circ}$, MY~24), after a lengthy
period of aerobraking. Only about six months of global observations are
available in MY~23 during the aerobraking, when the spacecraft orbit was still
very elliptical (6 hour orbit instead of the nominal 2 hour mapping phase
orbit). MGS stopped working properly in November 2006, and the mission was
officially ended in January 2007.

Mars Odyssey started its mapping phase in February 2002
($\mathrm{L_{s}}\sim330^{\circ}$, MY~25) and is still working at the time of writing. It is the longest running spacecraft orbiting
an extra-terrestrial planet to date.

Mars Reconnaissance Orbiter started its mapping phase in November 2006
($\mathrm{L_{s}}\sim128^{\circ}$, MY~28) and, although it encountered a few
problems that kept it in safe mode for an extended period at the end of 2009, is currently operating nominally. 

The Mars Exploration Rovers ``Spirit'' and ``Opportunity'' started
their missions on Mars respectively on 4th January 2004 and 25th January 2004
($\mathrm{L_{s}}\sim328^{\circ}$ and $\mathrm{L_{s}}\sim339^{\circ}$, MY~26).
Spirit ceased communications with Earth in March 2010
($\mathrm{L_{s}}\sim67^{\circ}$, MY~30), whereas Opportunity is still
active on the surface of the planet. 

For building the dust climatology described in this paper, we have used CDOD retrievals and estimates obtained from the following
instruments observing at IR wavelengths:
\begin{itemize}
\item TES aboard MGS, from $\mathrm{L_{s}}=103.6^{\circ}$ in MY~24 to $\mathrm{L_{s}}=82.5^{\circ}$ in MY~27. After this date, the number and quality of TES observations rapidly
decreased (August 2004). CDOD retrievals \citep{smi04} are in absorption (scattering is not modeled) from nadir observations
at wavelengths centered around 1075~$\mathrm{cm^{-1}}$ (9.3~$\mu\mathrm{m}$). Local times are narrowly centered around 2pm at most latitudes, except when the orbit crosses high latitudes.
\item THEMIS aboard ODY, from $\mathrm{L_{s}}=0^{\circ}$ in MY~26 (we did not use observations taken at the end of MY~25) to $\mathrm{L_{s}}=360^{\circ}$ in MY~ 31. The instrument is still working at the time of writing. CDOD retrievals \citep{smi03,smi09} are in absorption from nadir observations at wavelengths centered around 1075~$\mathrm{cm^{-1}}$ (9.3~$\mu\mathrm{m}$) as for TES. Local times are between 2.30pm and 6pm at most equatorial and mid-latitudes (except for latitudes polewards of $|60^{\circ}|$).
\item MCS aboard MRO, from $\mathrm{L_{s}}=111.3^{\circ}$ in MY~28 to $\mathrm{L_{s}}=360^{\circ}$ in MY~31. MCS was switched on in September 2006 before the official beginning of MRO primary science phase, but had a long period in MY~28 (between 9th February and 14th June 2007) during which a mechanical problem preventing the use of its elevation actuator forced the team to keep it in limb-staring mode. The event precluded any nadir or off-nadir observation during this period, with the effect of limiting the vertical extension of the retrieved profiles (including those of extinction opacity) in the lower part of the atmosphere. Since 9th October 2007, off-nadir measurements with surface incidence angles between about $60^\circ$ and  $70^\circ$ have resumed with nearly every limb sequence, but for retrievals of aerosol extinction opacities only limb views are currently used. \citet{kle09} obtained profiles of dust extinction opacity from limb observations at wavelengths centered around 463~$\mathrm{cm^{-1}}$ (21.6~$\mu\mathrm{m}$) as standard MCS product. In this paper we have used estimates of CDOD from the limb observations, as detailed in Section~\ref{qualityMCS}. Local times are centered around 3am and 3pm at most latitudes, except when the orbit crosses the polar region. Since 13th September 2010 ($\mathrm{L_{s}}=146^{\circ}$, MY~30) MCS has also been able to observe cross-track, thus providing information in a range of local times at selected positions during MRO orbits \citep{kle13}. We include these cross-track observations in our gridding, when available.
\end{itemize} 

For the validation of the gridded maps with independent observations, we have used retrievals of near-IR and IR CDOD from the following instruments:
\begin{itemize}
\item PanCam cameras \citep{bel03} aboard MER-A ``Spirit'' and MER-B
``Opportunity'', respectively from $\mathrm{L_{s}}\sim328^{\circ}$ and
$\mathrm{L_{s}}\sim339^{\circ}$ in MY~26. Spirit PanCam stopped providing measurements
after $\mathrm{L_{s}}\sim67^{\circ}$ in MY~30. \citet{lem14} have retrieved CDOD from upward-looking nadir observations at wavelengths centred around 880 nm (near IR) and 440nm (visible blue). In this work we use only retrievals at 880 nm. There are no significant differences between the values provided at the two wavelengths.
\item Mini-TES aboard MER ``Spirit'' and ``Opportunity'' \citep{chr03}, starting from
the same times as PanCam cameras until $\mathrm{L_{s}}\sim191^{\circ}$ (Spirit)
and $\mathrm{L_{s}}\sim269^{\circ}$ (Opportunity) in MY~28. After $\mathrm{L_{s}}\sim269^{\circ}$ the detectors
were covered by dust from the MY~28 planet-encircling dust storm and became
unreliable. \citet{smi06} have retrieved CDOD from upward-looking nadir observations at IR wavelengths centered
around 1075~$\mathrm{cm^{-1}}$ (9.3~$\mu\mathrm{m}$).
\item CRISM \citep{mur07} aboard MRO, from $\mathrm{L_{s}}\sim133^{\circ}$ in
MY~28 until $\mathrm{L_{s}}=360^{\circ}$ in MY~30. CDOD retrievals have been obtained from nadir observations at
wavelengths centred around 900~nm \citep{wol09}. Although there are CRISM observations available after $\mathrm{L_{s}}=360^{\circ}$ in MY~30, and CRISM is still working at the time of writing, we are not using these observations in the present work because an issue of truncated EPFs in MY~31 requires further investigation.  
\end{itemize} 

Figure~\ref{fig_timeline} shows a summary of the time periods for which
observations from the above mentioned instruments are available, together with
the time limits within which we have used them in this paper.

\begin{center}--- Figure~\ref{fig_timeline} ---\end{center}

In addition to CDOD data, we have also used visible wide-angle pictures from
MGS/MOC \citep{mal10}, and visible pictures
from MRO/MARCI \citep{bel09}. The main purpose is to compare the evolution of selected dust storms in the reconstructed CDOD maps to the daily evolution that can be appreciated in camera images of the Martian surface. 

%--------
\subsection{Data quality control}
\label{quality}

Before using the CDOD datasets retrieved from the different instruments, we have
firstly checked the values against quality control criteria to eliminate
unreliable data. 

For all datasets, negative values are only allowed if the sum
of the value and the associated uncertainty (see Section~\ref{uncertainty}) is
non-negative. Although a single negative measurement (in this specific case a
`measurement' corresponds to a `retrieval') can be considered unphysical, the weight of negative measurements must be taken into account in statistical calculations to prevent biases towards statistically non-zero values. It is anticipated that both
the gridded values described in Section~\ref{gridding} and the interpolated ones
described in Section~\ref{kriging} are only accepted if they are positive,
otherwise they are replaced by a minimum positive value of 0.02. %explain why 0.02?

For specific datasets, we have adopted the following
quality control criteria. We show in Figures~\ref{fig_retrievalnumberday} and \ref{fig_retrievalnumbernight} the number of CDOD retrievals used in the present work.  

\subsubsection{TES and THEMIS}
\label{qualityTESTHEMIS}
For TES, we have retained CDOD observations flagged as ``good quality'' by the TES team, which passed additional quality control criteria. Specifically, we require that the surface
temperature is greater than 220~K \citep{smi04}, the difference between surface temperature and maximum atmospheric temperature is greater than 5 K, the radiance fit residual is
lower than 20, the opacity of the carbon dioxide hot bands is between -0.01 and 0.05, and
the water ice opacity is greater than -0.05 (the negative value threshold is
constrained by the uncertainty). The surface temperature threshold insures a
good signal-to-noise ratio, whereas the threshold on the temperature difference between surface and atmosphere prevents unreliable retrievals when the temperature contrast is too small. The latter case is particularly valid during high dust loading conditions, as showed by results from the Geophysical Fluid Dynamics Laboratory GCM (GFDL-GCM) and comparison with MOC images during the 2001 planet-encircling dust storm \citep{wil11}. With the choice of these quality control criteria, retrievals are practically limited to daytime conditions over ice-free surfaces, and may be reduced during intense dust storms (see Figs~\ref{fig_retrievalnumberday} and \ref{fig_retrievalnumbernight}). 

%\citet{mon13} shows a plot of the number of
%CDOD retrievals passing the quality control procedure described above, binned in
%$5^{\circ}$~solar longitude and $2^{\circ}$~latitude, for the four Martian years
%of TES mapping. TES observations are typically spaced at $30^{\circ}$~ longitude
%increment, with a drift in longitude on each subsequent day.

%Furthermore, we discarded a few observations at latitude higher than
%$42^{\circ}$ south in MY~26 belonging to three specific orbits between
%$\mathrm{L_{s}}\sim42^{\circ}$ and $\mathrm{L_{s}}\sim69^{\circ}$. The reason
%for this is that these observations occurred at latitudes much higher than the
%rest of observations in this period, and the values of dust opacity are
%surprisingly large.

THEMIS CDOD retrievals include several ``framelets'' in the same ``image'',
i.e. several values within the same stripe of observations. Only the last
framelet of each image is properly calibrated, but in this study we have
considered all available framelets to increase the number of THEMIS
observations, especially in MY~27 when we use no other observations. We
associate a slightly larger uncertainty to non-calibrated framelets, as
explained in Section~\ref{uncertainty}. For quality control purposes, we have retained CDOD values when the rms residual from
the aerosol opacity fit is lower than 0.4, and the surface temperature is greater
than 210~K.

\subsubsection{MCS}
\label{qualityMCS}
MCS observes the atmosphere in limb and off-nadir modes, which has the advantage
of allowing the retrieval of vertical profiles of dust opacity, but has the
disadvantage of not being able to observe the first few kilometres above the
ground (depending on the angle of observation). 
In order to estimate the column optical depth for aerosol dust, the MCS team has integrated the full profile of dust extinction opacity produced by the retrieval algorithm (v4.0 is the version used in this work) after a successful
retrieval. The profile is extended upward and downward under the assumption of
well mixed dust, based on the last valid value. The standard retrieved dust profile is often
truncated when regions have measured radiances that are not fit by the forward
radiative model within defined thresholds in dust, water ice or temperature. In the upper levels, the minimum extinction coefficient threshold for dust is $10^{-9}$ km$^{-1}$. In the lower levels, the profiles can get saturated by high opacity values, exceeding the threshold of $10^{-3}$ km$^{-1}$ \citep{kle09}.
Column optical depth estimates, on the other hand, use non-truncated profiles to make
sure all available and reasonable information is taken into account. The MCS dust
profiles cannot be retrieved down to the surface, and the dust in the
un-retrieved part of the profile can account for a significant fraction of the
total dust column. Estimates of CDOD from MCS observations, therefore, are
likely to introduce errors attributable to either the extrapolation to the
surface under the well mixed assumption or the use of dust opacity values at
altitudes where the fit to observed radiances is not within the standard
threshold. This problem may be particularly acute in light of evidence of
elevated dust layers \citep{hea11,guz13b}. 
 
MCS CDOD estimates can be fairly inaccurate if the lowest retrieved level
of dust opacity is above $\sim20$~km altitude, depending on the time of the year and
the dust/water ice conditions. We have therefore retained CDODs that correspond
to dust extinction profiles with valid values at least at or below 25 km
altitude. Dayside retrievals are particularly affected by low altitude
aereosols. To prevent large extrapolation of dust extinction coefficient to the
surface during daytime observations, we only accept estimated daytime CDODs when
the corresponding extinction profile has valid values below 8 km altitude.

Furthermore, we have rejected CDOD estimates when the temperature profile
dropped below the condensation temperature of carbon dioxide at some pressure
levels, because CO$_2$ ice opacity can affect retrievals of dust opacity at
those levels (carbon dioxide ice opacity is not currently taken into account in
MCS retrievals). It is worth mentioning that this selection might introduce a
systematic bias, especially during the northern spring and summer seasons in the
tropics. TES provides daytime CDODs while MCS estimates of CDOD in this season are largely based
on nighttime viewing, because daytime clouds in the aphelion tropical belt limit
the success of retrievals.   

%We have eliminated a very large and isolated peak at sol 217 in MY~28 from the Spirit time series, which is possibly due to the occurrence of an external event.
%It is important to remind here that the values of dust opacity observed by MER-Opportunity PanCam camera in northern spring-summer are somehow higher than what every other instrument measures. It has not been possible so far to establish whether these observations are physically reliable or a systematic problem affects Opportunity observations at that time of the year. For this reason and for the purposes of building dust scenarios, we used the minimum sol-averaged dust opacity observed by Spirit and Opportunity for each sol.   

\begin{center}--- Figure~\ref{fig_retrievalnumberday} ---\end{center}

\begin{center}--- Figure~\ref{fig_retrievalnumbernight} ---\end{center}

%--------
\subsection{Data processing}
\label{dproc}

The CDODs from TES/THEMIS and MCS are retrieved at different IR
wavelengths, in absorption for TES/THEMIS and in extinction for MCS,
which prevents the direct comparison. We account for this problem by converting
the estimated MCS extinction CDODs at 21.6~$\mu\mathrm{m}$ into equivalent
absorption CDODs at 9.3~$\mu\mathrm{m}$. We carry out the conversion by
multiplying the MCS values by a factor 2.7,
which takes into account the effective dust particle radius
(1.06~$\mu\mathrm{m}$) and variance (0.3) used in the retrievals \citep[version
4.0, ][]{kle09,kle11}. It is worth mentioning that the dust size distribution is maintained fixed throughtout all seasons and locations. 

We use CDOD from PanCam cameras aboard MER-A and MER-B rovers
at the near-IR wavelength of 880 nm.
CRISM CDOD retrievals are obtained at the near-IR wavelength of 900 nm. When
using PanCam and CRISM for validating the gridded maps in
Section~\ref{validation}, we have to account for the differences between the
absorption-only gridded values at 9.3~$\mu\mathrm{m}$ and the full extinction
near-IR retrievals. This is done by converting the gridded values into
equivalent visible values. The process firstly converts values from
absorption-only at 9.3~$\mu\mathrm{m}$ to full-extinction values by multiplying
by a factor 1.3 \citep{smi04,wol03}, then converts the latter to mean visible
values by multiplying by a factor 2.0, consistent with 1.5- to
2.0-$\mu\mathrm{m}$ dust particles \citep{cla03,lem04,wol06}. Overall, the
factor we use to convert from absorption-only at 9.3~$\mu\mathrm{m}$ to equivalent full-extinction visible is 2.6. This factor, though, is
affected by large errors, deriving from both the conversion to extinction
\citep{smi04}, and the infrared-to-visible conversion \citep{lem04,wol06}, see
also Section~\ref{uncertainty}. This is the reason why we have decided to
provide our gridded products at the original IR wavelengths, despite the
advantage of producing equivalent visible CDOD maps for global and
mesoscale atmospheric models, whose radiation schemes usually compute dust
heating rates based on mean visible opacities using assumed IR/visible ratios. 

In order to calculate spatial averages, we require that CDODs are
normalized to a reference pressure level, hence eliminating the effect of
topographic inhomogeneities. The retrieved values are divided by the surface
pressure at the appropriate location and time of year, extracted from the pres0
tool included in the MCD v4.3 \citep{mil11}, and multiplied by 610 Pa, which is
our choice for the reference pressure level. The CDODs used
in the present study, therefore, are equivalent to integrate or extrapolate the
optical depth down to 610 Pa at all locations on the planet, under the assumption of well-mixed dust. See
Section~\ref{uncertainty} for a discussion of uncertainties associated to
surface pressure.  

%--------
\subsection{Data uncertainties}
\label{uncertainty}

We have estimated the uncertainty of each single CDOD value that we
have used in this work. The total uncertainty of an observation is provided as
the propagation of single uncertainties detailed in this Section, for each
instrument. The error propagation is carried out using relative errors, which
highly simplify the formulas. The squared relative propagated error reduces to
the sum of squared relative errors on each independent variable when operations
among variables include only multiplications and divisions, as in our case. 

An uncertainty that is common to all observations is associated to the value of
the surface pressure extracted from the MCD v4.3, which is used to extrapolate CDODs to a specific pressure level. After the corrections for  topography and total mass of the atmosphere, which are carried out
within the pres0 routine of the MCD, the largest sources of surface pressure
variability on Mars are the weather systems, mainly in the form of dust storms and high latitude
winter baroclinic waves. In order to account for this
variability, we have extracted the day-to-day RMS of surface pressure from the
MCD \citep[see the MCD Detailed Design Document associated to the version of the
database described in ][]{mil11} to build a $5^\circ$ solar longitude
$\times5^\circ$ latitude array. We provide an estimate of the surface pressure
uncertainty associated to the specific location and season of an observation by
interpolating this array.

\subsubsection{TES and THEMIS}
Retrievals of CDOD from TES and THEMIS observations come with an
associated nominal uncertainty, estimated by taking into account random errors
in the instrument and calibration, as well as possible systematic errors in the
retrieval algorithms. \citet{smi04} gives an estimate for the total uncertainty
in TES absorption-only IR dust optical depth for a single retrieval of 0.05 or
10\% of the column optical depth, whichever is larger. \citet{smi09} provides
this value for THEMIS absorption-only IR dust optical depth for the calibrated
framelet: 0.04 or 10\% of the column optical depth, whichever is larger.

For THEMIS observations, \citet{smi09} also states that ``uncertainties are
likely somewhat higher (perhaps 20\% or even higher) during the most intense
dust storms because large corrections to the temperature profile must be made
for those observations''. It is likely that TES retrievals during high dust
loading are also affected by larger uncertainties, due to the reduced
surface-atmosphere temperature contrast and uncertainty in dust size
distribution. In order to account for this, for TES we use the nominal
uncertainty up to IR optical depth 1.0, then we increase the uncertainty to 20\%
up to IR optical depth 2.0, and to 30\% for IR optical depths greater than 2.0.
For THEMIS, we use the nominal uncertainty up to IR optical depth 0.5, we
increase the uncertainty to 20\% up to IR optical depth 2.0, and to 30\% for IR
optical depths greater than 2.0.

As for the non-calibrated THEMIS framelets, we currently increase their
uncertainties by 20\% (e.g. 0.06 instead of 0.05 when the CDOD value is 0.5)

\subsubsection{MCS}
As explained in Section~\ref{quality}, the procedure adopted to estimate CDOD from MCS observations is likely to introduce errors, particularly when the
truncated dust profiles do not have valid values close to the surface. 
Estimating these errors is particularly difficult, as crucial information may
be missing in the atmospheric layers where the dust loading is significant. Our
estimate of the uncertainty on MCS CDODs, therefore, can
only be empirical. In Section~\ref{summary}, we mention future developments
in relation to MCS CDOD retrievals.

In this work, we assign an uncertainty on the basis of the altitude of the
lowest valid level in the dust profile used for the vertical integration. The uncertainty is
linearly increasing from 5\% (nominal value) to 60\% as a function of this
altitude, from the surface to the highest accepted level of 25 km. Although this
uncertainty estimate is arbitrary, it provides a way to distinguish MCS values
that are likely to be biased, and to relatively evaluate this bias. This
feature is used during the gridding procedure, as detailed in
Section~\ref{gridding}.    

Very small values of CDOD estimated from dust extinction profiles that do
not have valid values below 4 km are considered spurious and substituted with
the minimum value of 0.01, with 10\% uncertainty (i.e. 0.001).

Finally, we consider that the 2.7 factor used to convert MCS extinction CDODs at
21.6~$\mu\mathrm{m}$ to equivalent absorption CDODs at 9.3~$\mu\mathrm{m}$ is
affected by 10\% uncertainty (i.e. 0.3), due to the uncertainty on the particle sizes. We take it into account in the
propagation of uncertainties for MCS.

\subsubsection{PanCam and Mini-TES}
We have averaged CDOD retrievals from PanCam and Mini-TES observations in each
available sol for both MER-Spirit and MER-Opportunity. We have also interpolated
in time to fill the gaps in the sol-averaged time series for both MERs. For sol-averaged PanCam and Mini-TES CDODs, we use the largest value between the standard deviation and the average of single measurement errors as a measure of uncertainty on the daily mean. This choice takes into account CDOD variations with local time as well as single measurement errors.

\subsubsection{Digression on absorption/extinction and IR/visible conversions}
As explained in Section~\ref{dproc}, we have decided to work at the original TES
and THEMIS absorption IR wavelength of 9.3~$\mu\mathrm{m}$, except when
comparing with PanCam and CRISM retrievals. The conversion factors from
absorption to extinction and from IR to visible depend on the aerosol refractive
indexes, which ultimately depend on the aerosol size, shape and composition, and
are likely to vary with seasons and locations. Both factors are therefore
affected by large uncertainties. \citet{smi04} shows in Fig.~4a a graph of the
relationship between effective absorption and full extinction optical depth for
dust at the equator at northern autumn equinox in MY~24 (it is stated that the
curve is typical of other times and locations). For IR dust optical depths lower
than 0.5, the relationship is fairly linear ($\tau_{abs}=\tau_{ext}/1.3$). The
departure from the linear relationship at higher dust optical depth can be
estimated as $-{\tau_{ext}}^2/25$. This estimate could be used to introduce an
uncertainty on the 
absorption-to-extinction factor, which increases with the optical depth. As for
the infrared-to-visible factor, uncertainties have been estimated in a few
measurement campaigns, although limited in time and location. \citet{lem04}
reports a value of $2.0\pm0.2$ from MER-PanCam observations, \citet{wol06}
reports in Table~2 dust optical depth values from MER-PanCam and Mini-TES, which
can be averaged to provide a value of $2.5\pm0.6$.

In Section~\ref{validation}, we have simply used a single factor
2.6 ($1.3\times2.0$) throughout the time series to convert the gridded values to equivalent visible values,
without explicitly taking the uncertainty on this factor into account.

%%%%%%%%%%%%%%%%%%%%%%%%%%
\section{Gridding dust opacity observations}
%%%%%%%%%%%%%%%%%%%%%%%%%%
\label{gridding}

The process of creating uniformly-spaced data (i.e. a regular grid) from
irregularly-spaced (scattered) data is generally known as `gridding'. There
exist several techniques to solve this problem, depending on the applications. A
basic technique commonly adopted when dealing with orbital spacecraft
observations is `binning' using space-time box averages. The original data
values which fall in a given multi-dimensional interval (a `bin') are replaced
by a value representative of that interval, often the average, but the median
can also be used to filter outliers.  If the bins are uniformly distributed, the
problem of gridding the data is solved. For our purposes, the simple box-average
binning is not suitable because 1) we desire to achieve the highest possible
spatio-temporal resolution for the CDOD maps, compatibly with having a
reasonable number of observations to grid, and 2) simple box averages introduce
temporal and spatial biases if long time intervals and/or large spatial boxes
are used (several orbits at different times are inevitably averaged together).  

Reconstructing the dust climatology while preserving the desired variability of dust storms at short time
scale and small spatial scale, therefore, requires a
more sophisticated data gridding. In order to achieve this objective, we
have developed an efficient `iterative weighted binning' methodology, described in the
following Section.
  
%--------
\subsection{The principle of iterative weighted binning (IWB)}
%--------
\label{bintheory}

The application of our gridding procedure is equivalent to a moving
weighted average characterized by the use of successive spatio-temporal windows.
The basic principle of the methodology is that, for a given grid point
at a given time, all observations within a defined time window and spatial range
are averaged using weights that depend on 1) the lag between the time of each
observation and the required time at the grid point, 2) the spatial distance
between each observation location and the location of the grid point, and 3) the
uncertainty (standard deviation) of each observation. 
In this work, we use the
Mars Universal Time (MUT) of the prime meridian as absolute time variable to
which we refer both the time of observations and the time of grid points on a synoptic
map.

The weighted average value $Y$ at a given time $t_0$ and location $\mathbf{x}_0$
for a generic observed variable $y$, which is a function of time and spatial
coordinates, is given by:
\begin{equation}
\label{eq1}
 Y(\mathbf{x}_{0},t_{0})=\frac{\sum\limits_{n=1,N} M(\delta\mathbf{x}_{n},\delta
t_{n})\cdot R(\delta t_{n})\cdot Q(\sigma_{y_{n}} / y_{n})\cdot
y_{n}(\mathbf{x}_{n},t_{n})}{\sum\limits_{n=1,N} M(\delta\mathbf{x}_{n},\delta
t_{n})\cdot R(\delta t_{n})\cdot Q(\sigma_{y_{n}} / y_{n})},
\end{equation}
where $n$ represents the index of the $N$ observations $y_1,y_2,\dots y_N$ that
are included in defined time window and space range (a multi-dimensional bin built
around the time $t_0$ and the location $\mathbf{x}_{0}$). $R$ is the time weighting
function, which determines the contribution of the $n^{th}$ observation according
to the time difference $\delta t_{n}=t_{n}-t_0$ (positive for observations in
the future, negative for observations in the past). $M$ is the distance weighting
function, which determines the contribution of the $n^{th}$ observation according
to its distance $\delta\mathbf{x}_{n}=|\mathbf{x}_{n}-\mathbf{x}_{0}|$ from the
location $\mathbf{x}_{0}$. The distance weight is also a function of the time
lag $\delta t_{n}$ so that observations at different times in the past or in the
future with respect to $t_0$ have different distance weights. $Q$ is the
uncertainty weight, which determines the contribution of the $n^{th}$ observation
according to its relative uncertainty ($\sigma_{y_{n}}$ is the observation 
standard deviation).

The particular choice for the form of the weighting functions $M$, $R$, and $Q$ depends on the specific application. For our purpose
of creating synoptic maps of CDOD using orbiting satellite
observations, it is convenient to use weighting functions similar to those implemented in the
analysis correction data assimilation scheme of the UK Meteorological Office
\citep{lor91} and in the derived Mars analysis correction data assimilation
scheme \citep{lew07}. In fact, Eq.~\ref{eq1} applied to $k$ grid points of a
regular grid can be considered as the weighted average equivalent of the model
grid point increment Eq.~3.18 in the analysis correction scheme of
\citet{lor91}, although this analogy has no formal basis.

Eq.~\ref{eq1} applies to observations in specifically defined time window and space range. Our methodology includes the iterative use of less and less restrictive time windows and space ranges, as we explain in the next section where we describe the application of IWB to the problem of gridding dust opacity observations.

%--------
\subsection{Gridding with IWB: procedure}
%--------
\label{bingrid}

If we apply Eq.~\ref{eq1} to each grid point $k$ of a pre-defined space-time
grid for the CDOD variable $\tau$, the weighted average
value $T$ at each grid point can be written as:
\begin{equation}
\label{eq2}
T_{k}=\frac{\sum\limits_{n=1,N} M_{nk}(\delta t_{nk})\cdot R(\delta t_{nk})\cdot
Q_{n}\cdot \tau_{n}}{\sum\limits_{n=1,N} M_{nk}(\delta t_{nk})\cdot R(\delta
t_{nk})\cdot Q_{n}},
\end{equation}
where $\tau_1,\tau_2,\dots \tau_N$ are the CDOD observations that are included in
a specified time window and distance range built around the time and location
of the grid point $k$. 
Additionally, one can define the weighted root mean squared deviation (or
weighted standard deviation) associated to the weighted average value $T_{k}$
as:
\begin{equation}
\label{eq2bis}
\sigma_{T_{k}}=\sqrt{\frac{\sum\limits_{n=1,N} M_{nk}\cdot R\cdot Q_{n}
(\tau_{n}-T_{k})^2}{\sum\limits_{n=1,N} M_{nk}\cdot R\cdot
Q_{n}}}=\sqrt{\frac{\sum\limits_{n=1,N} M_{nk}\cdot R\cdot Q_{n}
\cdot\tau_{n}^2}{\sum\limits_{n=1,N} M_{nk}\cdot R\cdot Q_{n}}-T_{k}^2},
\end{equation}

We choose $M$ to be a second-order autoregressive correlation function of the
distance $d_{nk}$ from observation $n$ to grid point $k$ and of the correlation
scale $S$ \citep[see also Eq.~3.19 in ][]{lor91}:
\begin{equation}
\label{eq3}
M_{nk}=(1+d_{nk}/S(\delta t_{nk}))\exp(-d_{nk}/S(\delta t_{nk})),
\end{equation}
where $S(\delta t_{nk})$ is a function of the time difference between
observation time and grid point time. $S$ is chosen to be a linearly increasing
function of this difference, symmetric with respect to $\delta t_{nk}=0$, with a
minimum value $S_{min}$ at $\delta t_{nk}=0$ and a maximum value $S_{max}$ at
the extrema of the chosen time window (TW). The equation for $S$ (defined over
the range $-\textrm{TM}/2<\delta t<\textrm{TM}/2$) is therefore provided by:
\begin{equation}
\label{eq3bis}
S=\frac{S_{max}-S_{min}}{TW/2}|\delta t|+S_{min}.
\end{equation}
With this choice of $S$, observations with times closer to the time of the
requested grid point have weights $M_{nk}$ that decrease faster with distance
(see Fig.~\ref{fig_weights}). In other words, smaller weights are assigned to
distant observations when their time is closer to the current time. A further
parameter, $d_{cutoff}$, sets the distance range on the spherical surface within
which the contributions of the observations to the average are considered. We
use the haversine formula to calculate the distance between two locations on the
spherical planetary surface, which is numerically better conditioned for small
distances \citep{sin84}.

%\begin{figure}[H]
% \noindent\includegraphics[width=0.65\textwidth,angle=-90]{fig_mu}
%  \caption{Plot of the distance weight function $\mu$ as a function of $d$ (see Eq.\ref{eq3}) for two different values of correlation scale, corresponding to $S_{min}=150$~km and $S_{max}=300$~km. The corresponding $S$ function is plotted in the right panel of this Figure, as a function of $\delta t$, defined over the range $-\textrm{TM}/2<\delta t<\textrm{TM}/2$. The cut-off distance $d_{cutoff}$ is in this case located in the left panel at 800 km.}\label{fig_mu}
%\end{figure}

$R$ is chosen to be a decreasing quadratic function of the difference $\delta
t_{nk}$ between observation time and grid point time \citep{lor91}, symmetric
with respect to $\delta t_{nk}=0$. 
%The justification for an asymmetric function in the analysis correction data assimilation scheme comes from the adjustment time of the model to the insertion of observations, but in this case there is no model adjustment involved. 
The equation for $R$ (defined over the range
$-\textrm{TM}/2<\delta t<\textrm{TM}/2$) is therefore:
\begin{equation}
\label{eq3tris}
R=\left(\frac{R_{min}-1}{TW/2}|\delta t| + 1\right)^2,
\end{equation}
where $R_{min}$ is the minimum value at the extrema of the chosen time window
(see Fig.~\ref{fig_weights} for an example with $\textrm{TW}=7$).

%\begin{figure}[H]
%  \noindent\includegraphics[width=0.65\textwidth,angle=-90]{fig_r2}
%  \caption{Plot of the quadratic time weight function $R^2$ for a symmetric time window of 11 sols. The value of the $R$ function at the extrema of the time window interval (-5.5 and +5.5 sols in this case) is set to 0.05. In this work, we always use this value for the parameter $R_{min}$, except when the iterative weighted binning is performed with a time window of 1 sol, in which case $R_{min}=1$ and the $R^2$ function assumes the constant value of 1.}\label{fig_r2}
%\end{figure}

Finally, $Q$ is chosen as a second-order autoregressive correlation function of
the relative uncertainty of observation, namely: 
\begin{equation}
\label{eq4}
Q_{n}=\left(1+\lambda\frac{\sigma_{\tau_{n}}}{\tau_{n}}
\right)\exp\left(-\lambda\frac{\sigma_{\tau_{n}}}{\tau_{n}}\right),
\end{equation}
where $\sigma_{\tau_{n}}$ is the uncertainty (standard deviation) of each
observation, and $\lambda$ is a scaling factor used to obtain the desired width
at half maximum (WHM) for the $Q$ function. We choose $\lambda=8.39173$ to have
WHM=0.2 (i.e. $Q=0.5$ when the relative uncertainty of an observation is
$20\%$). See Fig.~\ref{fig_weights} for a plot of the $Q$ function we use. 

%\begin{figure}[H]
%  \noindent\includegraphics[width=0.65\textwidth,angle=-90]{fig_q}
%  \caption{Plot of the relative error weight function $Q$ (see Eq.~\ref{eq4}). The scaling factor $\lambda$ is set to 5 in this work. The reliability value $\varepsilon$ is greater than zero and varies mostly between 0 (reliable observation) and 1 (unreliable observation). The $Q$ weight is 0.03 when $\varepsilon=1$.}\label{fig_q}
%\end{figure}

\begin{center}--- Figure~\ref{fig_weights} ---\end{center}

The following is a summary of our gridding procedure with IWB.

\begin{enumerate}
\item We have tested the sensitivity of the method to different spatial and
temporal resolutions and the results are reported in Section~\ref{bingridex}. On
the basis of the sensitivity tests, we have chosen the resolutions of the
longitude-latitude grids reported in Table~\ref{tab1}, according to different
datasets. The choice of the time resolution is one sol. Each synoptic gridded
map is centred around local time noon at the prime meridian, i.e. a sol at
$0^{\circ}$ longitude is defined between 00:00 and 24:00 MUT. 
\item For MY~24 and 25 we have used only TES CDODs. For MY~26 and 27 we
have used both TES and THEMIS observations until $\mathrm{L_{s}}\sim 80^{\circ}$
in MY~27, then only THEMIS observations. For MY~28, we have used only THEMIS
until $\mathrm{L_{s}}\sim 112^{\circ}$, then both MCS and THEMIS, as well as in
most of MY~29, 30, and 31, apart from $\mathrm{L_{s}}\sim 327^{\circ}$, MY~29,
to $\mathrm{L_{s}}\sim 24^{\circ}$, MY~30, when only THEMIS observations are
available.
\item For each set of observations (TES, TES+THEMIS, THEMIS, MCS+THEMIS) we have
defined the parameters required by Eq.~\ref{eq2}, which are
summarized in Table~\ref{tab1}. The IWB procedure then requires that, for
each sol and for each spatial grid point, we define a time window and distance
range within which we use observations to calculate the weighted average and its
associated weighted standard deviation.
\item We repeat the previous step iteratively, using different time windows,
from the smallest to the largest (the values we use are reported in
Table~\ref{tab1}), and calculating the weights accordingly. At each iteration, more valid grid points are added to the synoptic map,
because more observations are considered, but we do not overwrite grid points
flagged as valid in previous iterations.
\item  The criterion to accept a value of weighted average at a particular grid point is
that there must be at least a minimum number of observations $N_{thr}$ within a distance
$d_{thr}$ from the grid point, having relative uncertainties lower than a
threshold value $(\sigma_{\tau}/\tau)_{thr}$ (see values in Table~\ref{tab1}).
If this is not the case, a missing value is assigned to the grid point at that
iteration.
\item The final result of the successive application of the weighted binning
equation consists in daily synoptic maps of CDOD
on the pre-defined regular grid, with missing values in places where the
gridding criterion was not satisfied at any iteration. 
\item  For the specific years MY~27 to 31, we apply an additional step in the procedure by substituting some of the missing values with valid gridded values obtained with only THEMIS observations using the parameters listed in Table~\ref{tab1}. This step is quite effective in reducing the gaps induced by the poor MCS coverage of major dust events and of the tropical regions affected by the aphelion cloud belt. It provides full weight to THEMIS observations with less restrictive gridding parameters, allowing for valid grid points which would otherwise not satisfy the acceptance criterion of the normal procedure.
\end{enumerate} 
 
\begin{center}--- Table~\ref{tab1} ---\end{center}

%--------
\subsection{Gridding with IWB: examples}
%--------
\label{bingridex}

In order to illustrate the procedure of gridding when the observation coverage
is good both in space and time (at least for most latitudes), we use an example
from TES observations in the northern winter of MY~24.

Fig.~\ref{fig_tes_time} highlights one of the main reasons why the application of
weighted binning in space and time is essential when gridding at high
spatio-temporal resolution. In this figure, TES retrievals of CDOD are
shown within a time window of 7 sols (about $5^{\circ}$ solar longitude at this season)
centered around noon at $0^{\circ}$ longitude in sol 449, $\mathrm{L_{s}}=227.8^{\circ}$, MY~24.
The area shown in the figure is limited to $80^\circ$ longitude and $40^\circ$
latitude south of the Equator. The colours indicate the time lag between the
time of an observation and the center time. It is clear that adjacent orbits
have quite different time lags. A simple box average would certainly introduce
biases in the creation of a synoptic map of CDOD for this sol. 

\begin{center}--- Figure~\ref{fig_tes_time} ---\end{center}

The result of the IWB with increasing
time windows is shown in Fig.~\ref{fig_tes_tw}. We separate each iteration (TW=1, 3, 5, and 7 sols) in the panel series (a) and (b), respectively showing the retrieved non-uniform values of TES CDOD, and the uniformly gridded values. Clearly, gridded values of CDOD with larger TWs are smoothed with respect to smaller TWs, even if time weighting is applied in both cases.  
The result of the successive application of weighted binning is then shown in panel (c), where the valid grid points in each
iteration add to the valid values of the previous one, without
overwriting. Eventually, we obtain a fairly complete, regular map of
CDOD where the observations closer to the MUT time of the synoptic map provide most of the unbiased information. This shows that an iterative weighted binning produces more valid and less biased values than the application of a single weighting with a fixed TW.

\begin{center}--- Figure~\ref{fig_tes_tw} ---\end{center}

The optimal spatial resolution for the gridded maps can be considered as the
highest one which includes a reasonable number of observations in each
spatio-temporal bin and, at the same time, produces results not dissimilar to those obtained
at a lower resolution. Ideally, the gridding method should not show much
sensitivity with respect to the choice of reasonable spatial and temporal
resolutions, particularly in the standard deviation field. In order to verify the quality of the choice we have made for the  datasets reported in
Table~\ref{tab1}, we have carried out a sensitivity test using four
different cases, as illustrated in Fig.\ref{fig_my24_restests} for a typical map
with high dust optical depth contrasts (same sol as in Fig.~\ref{fig_tes_tw}). Both averaged values and standard
deviations (which in this case represent the variability
within the group of averaged observations) show little variation throughout the
four cases. This result illustrates (at least for TES) that our gridding procedure is not
particularly sensitive to the choice of the 
spatial resolution within a limited range. This allows us to choose for TES in MY~24, 25 and 26 fairly high longitude and latitude resolutions, namely $6^\circ\times3^\circ$. The same latitude resolution cannot be used for MCS, given the fact
that the path of the limb observations usually spans several degrees in
latitude, and there are fewer observations per degree than TES. Finally, the
sparse distribution of THEMIS observations must be taken into account in the
choice of the spatial resolution, when only THEMIS data are available. In order to maintain a consistent resolution throughout MY~28, 29, 30 and 31 we use $6^\circ\times5^\circ$ in longitude and latitude.

\begin{center}--- Figure~\ref{fig_my24_restests} ---\end{center}

The combination of Fig.~\ref{fig_tes_tw}, \ref{fig_my24flush_evol}, and
\ref{MOC_TES_MY24} in Section~\ref{camval} provides the basis for our choice of
the 1-sol time resolution. Ideally, we want the highest time resolution to
follow the development of regional dust storms, using as many independent
observations as possible. Given the number of TES and MCS orbits per sol
(respectively 12 and 13), there are not enough independent observations to
provide a sufficient spatial resolution below the 1-sol time resolution. The top panels of the (a) and (b) series in Fig.~\ref{fig_tes_tw} show that the main features of the regional storm
occurring in MY~24 at $\mathrm{L_{s}}=227.8^{\circ}$ can be captured with a
1-sol TW gridding. Fig.~\ref{fig_my24flush_evol} illustrates that, for the same
period, we can characterize the daily evolution of the storm, and
Fig.~\ref{MOC_TES_MY24} demonstrates that such daily evolution can be followed
in MOC synoptic visible images, thus validating our gridded maps in this
context. TES has, nonetheless, the 
advantage of a good spatial coverage even in condition of high aerosol loading,
which is often not the case for MCS. Fig.~\ref{fig_my29flush_evol} shows the daily evolution of another regional storm occurring in MY~29 at $\mathrm{L_{s}}\sim235^{\circ}$, which clearly highlights the large number of missing grid points. As previously mentioned in Section~\ref{qualityMCS}, MCS CDOD estimates are few in numbers during episodes of high dust loading, and have large associated uncertainties because the retrieved dust extinction profile is missing in the dusty part of the atmosphere. The criterion we use for the acceptance of grid points, based on a minimum number of observations within a threshold distance having relative uncertainty below a defined threshold, is particularly strict for MCS and THEMIS observations. The thresholds we have used in this work (see Table~\ref{tab1}) might be relaxed from MY~28 onwards, especially the relative uncertainty threshold. Other ways to increase the number of valid grid points are discussed in Sections~\ref{camval} and \ref{summary}, and include the 
possibilities to estimate the CDODs from observations of brightness temperatures, using a GCM, and to retrieve proper CDODs from nadir and off-nadir MCS observations.

\begin{center}--- Figure~\ref{fig_my24flush_evol} ---\end{center}

\begin{center}--- Figure~\ref{fig_my29flush_evol} ---\end{center}

%%%%%%%%%%%%%%%%%%%%%%%%%%
\section{Validation of gridded maps}
%%%%%%%%%%%%%%%%%%%%%%%%%%
\label{validation}

We have carried out a statistical validation of the incomplete maps of CDOD based on two approaches:
\begin{enumerate}
\item an internal validation comparing TES and MCS CDODs to gridded CDODs,
interpolated in the locations and at the times of the observations; 
\item an external validation using independent observations. These include the
values of visible CDOD retrieved from CRISM observations, the time series
of visible CDOD retrieved from PanCam cameras aboard Spirit and Opportunity,
the time series of IR CDOD retrieved from Mini-TES aboard the two MERs, and
some MOC and MARCI images during the evolution of regional storms.
\end{enumerate}

%--------
\subsection{Internal validation}
%--------
\label{intval}

For the statistical internal validation, we have interpolated the gridded maps
(linearly in time and bilinearly in space) at the location and time of each
observation, if a complete set of neighbours was available (i.e. four adjacent
spatial grid points at two consecutive sols). We have interpolated values for
both the gridded average field and the standard deviation one. 

The first test we have carried out is a simple correlation test.
% For TES and MCS
% datasets, we have plotted in Figs.~\ref{fig_correlation_grid_TES} and
% \ref{fig_correlation_grid_MCS} the scatter plots of the interpolated values
% versus the corresponding observed values, which in the ideal case should line up
% on a straight line with slope 1. 
% 
% \begin{center}--- Figure~\ref{fig_correlation_grid_TES} ---\end{center}
% 
% \begin{center}--- Figure~\ref{fig_correlation_grid_MCS} ---\end{center}
For all examined years, there is clearly a correlation between the observed values and the reconstructed ones, despite the uncertainties introduced by the linear interpolation. The linear correlation is very good in MY~25, 28, 29, 30 and 31, as the high values of the Pearson correlation coefficient suggest ($r\geq0.92$), and the data points indeed accumulate around a straight line with slope close to 1. For the other years, the Pearson correlation coefficient provides lower values ($r\sim0.85$), but still observed values and reconstructed ones appear visually correlated (not shown here).
%Note the  much lower values of CDOD observed by MCS in comparison to TES, which would clearly bias the reconstructed dust climatology if THEMIS observations were not conjointly used. Wired correlation in TES MY 24, 25??

For each available interpolated value, we have then calculated the standardized
mean difference (SMD) between this value and the observed value, which is weighted using
the combined standard deviation. Under the approximation that the observed value
and the interpolated value are independent (i.e. the covariance is neglected),
the variable we calculate is:
\begin{equation}
\label{eq5}
\beta=\frac{\tau_{\mathrm{int}}-\tau_{\mathrm{obs}}}{\sqrt{\sigma_{\mathrm{int}}
^{2}+\sigma_{\mathrm{obs}}^{2}}},
\end{equation}
which is equivalent to expressing the difference between the two values in terms
of their combined standard deviation. Strictly speaking, the interpolated value
is not independent of the observed value because the latter has been used to
calculate the former, so the value of $\beta$ might be underestimated.
Nonetheless, this variable is useful for estimating whether clear biases are
present in our gridded maps, or whether the differences are within statistical
limits. 

We display in Figs.~\ref{fig_histo_grid} the histograms of the SMD for the TES (MY~24, 25, 26) and MCS datasets (MY~28, 29, 30, 31). The difference in the number of values from year to year reflects the number of available retrievals, as shown in Figs.~\ref{fig_retrievalnumberday} and \ref{fig_retrievalnumbernight}. All histograms show that most of the values have $\mathrm{SMD}\leq|1|$ (the difference is much lower than the combined standard deviation) with $\sigma_{\mathrm{SMD}}<0.6$, and peak values are very close to zero. These general characteristics provide a sound internal validation from the statistical point of view. Other important factors to be considered are the shape of the histograms (i.e. skeweness and kurtosis), which can highlight possible biases. All distributions are leptokurtic (i.e. more sharply peaked than a Gaussian distribution), which indicates differences with respect to the combined standard deviation generally smaller than what expected by a random process. MCS histograms are very symmetric (small skeweness) therefore no particular bias is evident. TES histograms, on the other hand, are clearly right-skewed, which suggests that, in the gridded maps of MY~24, 25 and 26, there are more largely overestimated CDODs than largely underestimated ones. The asymmetry of the SMD distributions for the TES years is consistent with lower values of the correlation coefficient for MY~24 and 26. The origin of such bias is mainly related to overestimated gridded CDODs in correspondence to small observed ones.

\begin{center}--- Figure~\ref{fig_histo_grid} ---\end{center}

A third diagnostic we use to internally validate our gridded maps is the relative standard deviation (STD) of the grid points. This is expressed by the ratio between the weighted STD, calculated with Eq.~\ref{eq2bis} (a measure of the variability of CDOD at a grid point), and the weighted average calculated with Eq.~\ref{eq2}. Fig.~\ref{fig_percrelerr} shows that MY~24, 29, 30, and 31 have distributions peaked at values of relative STD lower than 10\%. Distributions in the other years peak at relative STD lower than 20\%, with MY~26 and 27 being the worst. All years have very right-skewed distributions, but the values of relative STD of the grid points do not exceed much the values of relative uncertainty of the single retrievals, which can be larger than 30\%, particularly in the MCS years. Fig.~\ref{fig_percrelerr}, therefore, suggests reasonable values for the relative STD of the gridded maps, and provides an indirect validation of the goodness of the chosen spatio-temporal resolution. If the 
resolution was too coarse, in fact, one would expect large variability at most grid points, which is not the case here.

\begin{center}--- Figure~\ref{fig_percrelerr} ---\end{center}

\subsection{Validation with independent observations}
%--------
\label{extval}

We used independent CRISM retrievals of visible CDOD in MY~28, 29, and 30 to
compare with equivalent visible gridded values interpolated in the positions and at times of CRISM observations. As explained in Section~\ref{dproc}, we have multiplied the gridded CDOD by a factor 2.6 to estimate equivalent visible extinction CDODs, and compare to CRISM values, retrieved at 900~nm wavelength. 

As in Section~\ref{intval}, we have calculated the Pearson correlation coefficients ($r=0.79$ in MY~28, $r=0.55$ in MY~29,and $r=0.47$ in MY~30),  
%correlation plots (Fig.~\ref{fig_correlation_grid_CRISM}) and 
and produced histograms of the SMD (Fig.~\ref{fig_histo_grid_CRISM}) for each available year. The comparison with CRISM data produces less correlated values and more biased histograms than the internal validation. In particular, the SMD distributions are clearly left-skewed, which suggests a majority of large underestimated values of CDOD in the gridded maps with respect to those observed by CRISM. It must be remembered, though, that uncertainties in the absorption-to-extinction and IR-to-visible factors would reflect in the skeweness of the SMD distributions. Nevertheless, only factors that combine to provide much larger values than the chose 2.6 value could make the distributions unskewed (not shown here). It is therefore very difficult to draw conclusions from the validation with CRISM observations in the visible.

%\begin{center}--- Figure~\ref{fig_correlation_grid_CRISM} ---\end{center}

\begin{center}--- Figure~\ref{fig_histo_grid_CRISM} ---\end{center}

We have also compared the gridded values, interpolated in the location of MER Spirit and Opportunity, to the CDOD measured by PanCam in the visible and Mini-TES in the IR on the two rovers. The time series of the comparison, for each sol when there are available observations, are shown in Fig.~\ref{fig_MERtimeseries}. We have chosen to display also MY~24, 25, and 26, even if MER observations are not available in those years, in order to increase the statistics on the interannual variability at Meridiani Planum and Gusev crater. As for the comparison with CRISM, we have multiplied the values extracted from the gridded maps and Mini-TES values by a factor 2.6, to convert them to a mean equivalent visible wavelength.

The comparison in Gusev crater is quite satisfactory throughtout the time series, with Mini-TES, PanCam and the gridded maps generally agreeing within the uncertainties (represented by the grey envelope), except for a couple of Mini-TES peaks not seen by PanCam as large. Even in the non-dusty seasons, the time series show consistency in all years, with values of CDOD between 0.2 and 0.3 on average.

The comparison in Meridiani is, on the contrary, satisfactory during the dusty season but problematic during the non-dusty one, when it highlights a strong difference between the CDOD measured from ground and the one gridded from satellite observations. Consistently in every year, the gridded maps provide values of CDOD that are about a half of those observed by PanCam and Mini-TES during the period $\mathrm{L_{s}}=[0^{\circ}, 180^{\circ}]$. This bias seems to be present in all three satellite datasets considered in this work, even when looking at single CDOD retrievals. \citet{lem14} have recently provided a possible explanation for this bias. According to them, water ice clouds and hazes contributed to the observed opacity at the Opportunity site in the summer season. Clouds were seen over the range $\mathrm{L_{s}}=[20^{\circ}, 136^{\circ}]$, with peak activity near $\mathrm{L_{s}}=50^{\circ}$ and $\mathrm{L_{s}}=115^{\circ}$. When looking at the Sun through the atmospheric column, PanCam in Meridiani is 
likely to add water ice optical depth to the dust optical depth, thus explaining the bias. \citet{lem14} confirms that ice clouds and hazes were not a significant part of the opacity at the Spirit site, which does not show biases with respect to observations from satellites.
Satellite measurements and gridded maps, therefore, should be considered more reliable in Meridiani at this time of the year. At other times, in fact, the differences between PanCam CDOD and gridded CDOD are well within the uncertainties, even during the MY~28 planet-encircling dust storm when only THEMIS observations are available for the gridding.     

When comparing the time series in Gusev and Meridiani, we can clearly observe and confirm from previous studies \citep[e.g.][]{vin09} that 1) the two planet-encircling dust storms (MY~25 and 28) had similar growth, peak values, and decay; 2) there are many regional storms that affected both Meridiani and Gusev (on the opposite side of Mars), reaching comparable peak values; 3) there is a tendency to have three distinct peaks of CDOD every second half of the year, with an early peak around $\mathrm{L_{s}}=180^{\circ}$, a major peak around $\mathrm{L_{s}}=240^{\circ}$, and a late peak around $\mathrm{L_{s}}=330^{\circ}$. The presence of three seasonal peaks of CDOD is not only confined to the equatorial latitude band, as it is discussed in Section~\ref{analysis} and shown in Fig.~\ref{fig_zonalmean_grid}.

\begin{center}--- Figure~\ref{fig_MERtimeseries} ---\end{center}

\subsection{Validation of dust storm evolution using camera images}
%--------
\label{camval}

We have used MOC and MARCI processed Daily Global Maps \citep{can01} to qualitatively compare in visible images
the evolution of the two selected regional storms shown in Section~\ref{bingridex}. Fig.~\ref{MOC_TES_MY24} shows the evolution of the
flushing storm occurring in MY~24 at $\mathrm{L_{s}}\sim227^{\circ}$, and 
Fig.~\ref{MARCI_MCS_MY29_3} shows the evolution of the regional storm occurring in MY~29 at $\mathrm{L_{s}}\sim235^{\circ}$.

Although the lack of strong contrast between the dust and the background in both MOC and MARCI images makes it difficult to clearly visualize the presence of dust, the overlaid satellite observations help to recognize where a storm is ongoing. In MY~24, the contours of gridded CDOD correlate remarkably well in position and shape with the presence of the dust haze in the MOC images at several sols during the evolution of the regional storm. This validation of the gridded maps during a particularly dynamic and fast-evolving event encourages their use for statistical analysis of the evolution of dust storms, particularly in MY~24, 25 and 26.

\begin{center}--- Figure~\ref{MOC_TES_MY24} ---\end{center}

Given the limited coverage of valid grid points in the maps of the MY~29 regional dust storm (see Fig.~\ref{fig_my29flush_evol}), it is not easy in this case to validate its evolution with the MARCI images. Nevertheless, it can be observed that the position of the dust haze in the MARCI images corresponds to the high values of CDOD in the gridded maps at corresponding sols. The overlaid satellite observations, in the case of the MARCI images, are surface temperature anomalies, which show a remarkable correlation with the dust haze. \citet{wil11} devised a technique to estimate the CDOD from satellite observations of surface temperature. Observed surface temperatures are actually top-of-atmosphere brightness temperatures in spectral regions where the atmosphere is relatively transparent. The technique uses a GCM with dust transport capability (the GFDL MGCM) to find the best value of dust optical depth at any given time and location that allows to match simulated brightness temperatures to observed ones. 
They applied this technique to estimate CDODs during the MY~25 planet-encircling dust storm using TES observations, and in principle the same technique can be applied to MCS observations of brightness temperature. Given the fact that brightness temperature observations are available even when proper CDOD retrievals fail, the gridding procedure would certainly benefit from the inclusions of CDOD data estimated as described above. 

\begin{center}--- Figure~\ref{MARCI_MCS_MY29_3} ---\end{center}

%%%%%%%%%%%%%%%%%%%%%%%%%%
\section{Multiannual dust climatology}
%%%%%%%%%%%%%%%%%%%%%%%%%%
\label{analysis}

The daily time series of CDOD spanning eight Martian years allows for analysis of intraseasonal and interannual variability. We have shown that the evolution of individual regional dust storms can be followed when the gridded maps are fairly complete. Even when the number of valid grid points in the daily maps is low, the CDOD variability can still be analysed statistically, provided data are averaged or filtered to prevent contamination from spurious values occurring at high frequencies (e.g. ``on/off'', isolated, large optical depths).

We provide an atlas of daily, irregularly gridded maps for each available Martian year as supplementary material of this paper. The reader can therefore explore the entire time series that forms the current version 1.5 of the CDOD climatology obtained with IWB, before downloading the corresponding NetCDF dataset from the LMD website (at the address reported in the footnote of Section~\ref{intro}). The atlas is organised as 669 pages (one for each sol) with eight maps on each page corresponding to the available Martian years. See also \ref{calendar} for a description of the sol-based Martian calendar devised in the present work. 

The structure of the atlas helps the analysis of interannual variability. It is interesting, for instance, to look simultaneously at the CDOD around sol 370 ($\mathrm{L_{s}}\sim179^{\circ}$) in MY~24, 25, and 26, a few sols before the beginning of the MY~25 planet-encircling dust storm. It appears clearly that the CDOD distribution was very similar in the three years observed by TES at this season, until sol 372 ($\mathrm{L_{s}}\sim180^{\circ}$), when slightly higher CDOD values appeared in the Hellas basin in MY~25, prior to expansion out of this crater towards the northern rim and in Hesperia Planum. Another feature that clearly stands out when looking at CDOD at high southern latitudes at this season is that TES observed fairly high values near the polar cap edge, where dust is likely to be lifted by southern baroclinic waves, active at this time of the year, whereas no sign of such high values is present in the other years. Very few THEMIS observations are available in MY~27 and 28, so it is difficult to 
judge these two years, but it is evident that MCS observations are biased 
towards very low CDOD values at high southern latitudes. 

This bias of MCS observations can be observed even more clearly in the zonal means of gridded CDODs, plotted as a function of solar longitude and latitude in each Martian year (Fig.~\ref{fig_zonalmean_grid}). There is an evident dichotomy between the TES years, where the high northern and southern latitudes in northern spring and summer present increased optical depth values, and the MCS years, where values decrease towards high latitudes. Panel (a) of Fig.~\ref{fig_stat_latallyears} offers a striking summary of this dichotomy around $\mathrm{L_{s}}=100^{\circ}$. The reason for this important bias is most likely the fact that dust is aloft only in the lower portion of the atmosphere (within the first kilometers) at this season and at these latitudes. MCS limb observations are not able to scan through these low levels, therefore dust is missed by the radiometer, which detects only very small values of opacity above the low dust layer. Future retrievals of CDOD from MCS in-planet observations, or estimates 
from top-of-atmosphere brightness temperature fit, might eventually correct this bias.

Biases apart, the latitudinal, seasonal, and interannual variability of CDOD can be fully appreciated in Fig.~\ref{fig_zonalmean_grid}, which summarizes eight years of dust climatology on Mars. This figure clearly highlights four distinctive phases in the distribution of dust during the second half of each year without a global-scale storm, and confirms what other studies have found, by using the longest available record of observations. Dust starts to increase in the atmosphere around northern autumn equinox in the southern hemisphere, but the largest increase usually occurs between $\mathrm{L_{s}}=220^{\circ}$ and $260^{\circ}$, when baroclinic activity at high northern latitudes favours cross-equatorial flushing storms (although not all regional storms occurring at this time originated in the northern plains). A third phase in the dust distribution is characterized by large lifting of dust occurring in the southern polar region between $\mathrm{L_{s}}=250^{\circ}$ and $300^{\circ}$, after the $\mathrm{CO_2}$ ice has mostly sublimated away. At other latitudes, instead, there is a clear decrease of atmospheric dust in every Martian year after $\mathrm{L_{s}}\sim260^{\circ}$ (with the exception of MY~28, characterized by the late planet-encircling storm). This pause in large dust storms coincides with the decrease in the amplitude of low-altitude northern baroclinic waves \citep[the so-called ``solsticial pause'', see e.g.][]{mul14}. When the solsticial pause is over, and the baroclinic wave activity at low altitude reinforces again, the probability of late flushing storms increases. Every year, therefore, a fourth phase in the dust distribution starts around or after $\mathrm{L_{s}}\sim320^{\circ}$, producing a late peak of CDOD.

\begin{center}--- Figure~\ref{fig_zonalmean_grid} ---\end{center}

The fairly repeatable pattern of CDOD from year to year contrasts with the highly unpredictable occurrence of global-scale dust storms, both in terms of frequency and season. If global-scale storms did not occur on Mars, and we did not take into account single regional storms, the ``typical'' annual distribution of CDOD would look like the one in Fig.~\ref{fig_dodtypyear}, where we have averaged all years together, excluding the largest CDOD value for each grid point. In this figure, the four distinct phases described above appear perfectly well defined. Obviously, also the MCS bias around the edges of the polar caps stands out in this figure.  

\begin{center}--- Figure~\ref{fig_dodtypyear} ---\end{center}

Figs.~\ref{fig_stat_eq_allyears} and \ref{fig_stat_latallyears} are useful to summarize similarities and differences among the years. At equatorial latitudes, there is little interannual variability in northern spring/summer, and low CDOD. The gradient of CDOD increases moving towards southern tropical latitudes and northern high latitudes in TES years, whereas in MCS years the bias mentioned above appears. In the second half of the year, the background dust level shows again little variability in the equatorial/tropical regions during the solsticial pause, except for the years with a global-scale storm. Large CDOD gradients are again observed moving towards high latitudes in TES years, but not in MCS years. Outside the solsticial pause, cross-equatorial flushing storms are likely to produce peaks of CDOD in the equatorial region, which affect all longitudes (as we have also seen in the comparison between Meridiani and Gusev).

\begin{center}--- Figure~\ref{fig_stat_eq_allyears} ---\end{center}

\begin{center}--- Figure~\ref{fig_stat_latallyears} ---\end{center}

As an example of how our gridded products can be used to explore the multiannual statistics of dust loading in particular regions, we consider the case of ESA's Exomars 2016 Entry, Descending and Landing Demostrator Module (EDM ``Schiaparelli''). The landing is planned to take place in Meridiani Planum during the time window $\mathrm{L_{s}}\sim240^{\circ}-250^{\circ}$ in MY~33, which is well inside the dust storm season. Although this will provide a unique opportunity to characterize a dust-loaded atmosphere during the entry, descending, and landing procedure, it will also pose strict contraints on the engineering parameters of the landing, and increase the associated risks. It is therefore very important to produce an accurate statistical prediction of the expected range of dust loading at the time and location of the landing, based on historical records. 
%%%%%In Fig.~\ref{fig_exomars}, we have plotted the time series of CDOD extracted from the gridded maps at Meridiani in the period $\mathrm{L_{s}}=225^{\circ}-265^{\circ}$ for all years, together with the cumulative histogram for the landing window $\mathrm{L_{s}}=240^{\circ}-250^{\circ}$. 
If one plots the time series of CDOD in all years in Meridiani, as in Fig.~\ref{fig_stat_eq_allyears}, but limited to the period $\mathrm{L_{s}}=225^{\circ}-265^{\circ}$, the tail of the planetary-encircling dust storm stands out with respect to all other years, which show moderate dust loading (not shown here). It is worth noting that this season is characterized by the solsticial pause of the baroclinic wave activity, with associated lack of flushing storms, which in general have trajectories with potential to affect Meridiani. By calculating the cumulative histogram for the landing window $\mathrm{L_{s}}=240^{\circ}-250^{\circ}$ (not shown here), we can make the statistical prediction, based on past observations, that Exomars 2016 ``Schiaparelli'' lander is likely to encounter a moderate dust loading with CDOD of $0.35\pm0.08$ (IR absorption referred to 610 Pa). This prediction, though, does not exclude the possibility of high dust loading induced by an equinoctial planet-encircling dust storm, which, as seen above, can be quite 
unpredictable even just a few sols beforehand. 

%\begin{center}--- Figure~\ref{fig_exomars} ---\end{center}

%%%%%%%%%%%%%%%%%%%%%%%%%%
\section{Building dust scenarios with kriging}
%%%%%%%%%%%%%%%%%%%%%%%%%%
\label{kriging}

With the application of the IWB procedure, the maps of CDOD that we obtain are incomplete and present higher spatial resolutions in MY~24, 25, 26 than in the other years. For many practical
applications it is desirable to have complete, regularly gridded maps spanning several years with the same resolution. One of such application is to prescribe realistic aerosol dust distributions for global-scale or
meso-scale climate model simulations. The production of the MCD statistics using the LMD GCM, for
instance, is an obvious example \citep{mil14}. Other possible applications include
retrieving surface or atmospheric variables (e.g. surface albedo or
atmospheric water vapor) using observations from which the
aerosol dust component needs to be subtracted. For these reasons, in this Section we discuss our
method to derive multiannual, regularly gridded `dust scenarios' from the irregularly gridded maps we have
described so far.

The process of producing complete gridded maps at a given resolution from maps
that have missing data and different resolutions is, generally speaking, a
problem of interpolation and/or extrapolation. There exist several techniques, more or less `optimal', to solve this problem, as it is the case for the gridding problem. 
 
Kriging\footnote{The word `kriging' is derived from the family name of Daniel G. Krige, whose
Master thesis the French mathematician Georges Matheron used to develop the
theory and formalism} (which is synonymous with `optimal
prediction') is a technique that belongs to the family of linear least squares
estimation algorithms. It is a method of interpolation that
predicts unknown values from data provided at known locations. Unlike other
common interpolation methods, such as polynomial, spline, and nearest-neighbor,
kriging does not require an exact fit at each tabulated data point. Another
important difference of kriging and other linear estimation methods is that
kriging aims to minimize the error variance of the predicted values. Kriging
applies a weighting to each of the tabulated data points based on spatial
variance and trends among the points. Weights are computed by combining
calculations of the spatial structure and dependence of the data, and building a statistical model of their spatial correlation (called `semivariogram' model). Alternatively, empirical semivariograms are often approximated by theoretical model functions, the most common of which are the spherical and the exponential semivariograms (we use the latter in this work).  The reader can refer to \citet{jou78} for a general overview of the method. For the application of kriging to data gridding, \citet{hay08} provide an example for the case of temperature ground stations on the Earth, and \citet{hof08} evaluate kriging among other methods of spatial interpolation. 

Given the spatial characteristics of the incomplete maps of CDOD that we have obtained after gridding the observations with the IWB procedure, kriging is the interpolation method that is likely to provide the best results, producing smooth spatial variations even in cases when many missing values are present.   

%--------
\subsection{Pre-kriging insertion of extra CDOD values}
%--------
\label{insertion}

Before applying the kriging to the maps, we have replaced some of
the missing values using the following methods (in order of application).
\begin{enumerate}
\item In TES, THEMIS and MCS datasets there are gaps in data coverage that
sometime extend for few sols, e.g. during Mars solar conjunctions. In those
cases, gridded maps are missing as well, if the gaps in data
coverage are longer than 7 sols (the maximum TW we use). In addition, one or two
daily maps before the first missing map and after the last missing map might have very few valid grid points in longitude. When long data gaps occur, we
have increased the TW up to 25 sols, in order to accumulate enough observations
to bypass the data coverage gap, and we have applied the gridding procedure
again with more successive iterations. This method produces a smooth time
interpolation by combining the moving average and the time weighting. In the
absence of alternative information, or of a dynamical model of dust transport,
this is a reasonable way to interpolate maps during periods of missing data. 
\item Issues of data coverage are both temporal and spatial. TES and
THEMIS datasets have few valid retrievals of CDOD at high
latitudes during the winter seasons, and the MCS dataset does not include
estimates when the dust loading is too high (e.g. during dust storms), the water
ice opacity is large (e.g. during the aphelion cloud tropical belt season), or
the temperature at some height is below the $\mathrm{CO_2}$ condensation
temperature (e.g. during the polar nights, when $\mathrm{CO_2}$ clouds might
form). In order to fill most of the spatial gaps in our gridded maps with climatological values, for each
sol of the Martian year we have used the average of all eight years, excluding from the average the largest value of CDOD to eliminate single dust storms.
%the median of all eight-year gridded maps, which automatically excludes all dust storms.
% the average of all eight-year gridded maps, excluding the periods of planet-encircling dust storm activity in MY~25 and 28.
For this climatological year (see Fig.~\ref{fig_dodtypyear} for a plot of its
zonal mean), spatial inhomogeneities are smoothed out, and the optical depths that result are
generally underestimated, particularly in the second half of the year. Before
using these 669 climatological maps to fill out gaps in the original gridded
maps, therefore, we have to re-normalize the values of CDOD using anchoring
values from independent observations. For each sol from MY~27 to 31 we
can use the observations of optical depth from PanCam `Spirit' and PanCam `Opportunity'
as anchoring value. We actually use the minimum between the sol-averaged
observations from Spirit and Opportunity, after having interpolated the time
series to fill any gap.  
%In MY~24--26, we use the minimum between the values of TES column dust opacity observations gridded with the IWB procedure at the locations of Spirit and Opportunity. 
The use of the minimum value avoids re-normalizing the climatological maps with
values characteristic of specific dust storms located at either Spirit's or
Opportunity's site. Furthermore, it is consistent with satellite observations in
northern summer (see Section~\ref{validation}), so that we avoid biases in the 
re-normalization. For a given sol, we calculate the average of the climatological CDOD map $\tau_{\mathrm{clim}}$ in a latitude band [$-15^{\circ}$,$0^{\circ}$] (we recall that the locations of Spirit and Opportunity are, respectively, close to $-14^{\circ}$ and $-2^{\circ}$ latitude), and we re-normalize the values of the map using the following factor $\nu$, weighted in latitude $\theta$:
\begin{subnumcases}{\label{eq6} \nu=}
 r_{\tau}+\frac{1-r_{\tau}}{2} \left( 1-\tanh\frac{\theta +45^{\circ}}{12^{\circ}} \right) & if  $\theta<0^{\circ}$ \\ r_{\tau}+\frac{1-r_{\tau}}{2} \left( 1+\tanh\frac{\theta -45^{\circ}}{12^{\circ}} \right) & if  $\theta>0^{\circ}$
\end{subnumcases}
where $r_{\tau}$ is the ratio between the minimum MER dust optical depth $\tau_{\mathrm{MER}}$ (converted to equivalent IR absorption by dividing by 2.6) and the calculated average $\tau_{\mathrm{clim}}$.
We replace a missing grid point in a gridded map from MY~27 to 31 with a value obtained as described above, if the
distance between the missing grid point and the closest valid grid point is
greater than a certain threshold distance, set equal to 1000 km. The threshold
avoids the introduction of possible artificial discontinuities between valid gridded
values and filled values in the maps.  
\item No valid grid points exist at very high latitudes during the polar
winters in TES and THEMIS years (MY~24 to 27). In order to
constrain the kriging interpolation at high latitudes and at the poles, we set
to 0.1 the value of the missing grid points located $20^{\circ}$ poleward of the
northermost or southermost valid latitude, including the value at the poles
(when missing). This value, although arbitrary, is suggested as being reasonable by
numerical simulations of the dust cycle in GCMs, and may be revised in future
updates of the dust scenarios, as new observations become available.
\end{enumerate}

%--------
\subsection{Interpolation by kriging}
%--------
\label{intkrig}

After replacing some missing values in the incomplete gridded maps, the data
fields are ready to be interpolated by krigging.

We have used the IDL built-in routine GRIDDATA for kriging interpolation on the sphere
with an exponential semivariogram model. We have chosen two different grids: a low resolution grid
($5^{\circ}\times5^{\circ}$ in longitude-latitude), and a high resolution grid
($2^{\circ}\times2^{\circ}$ in longitude-latitude). At the end of the
interpolation procedure, carried out for each available sol from MY 24 to MY 31,
we obtain complete, regularly gridded maps of IR absorption CDOD at the reference pressure of 610 Pa.
Fig.~\ref{fig_krigmap} shows an example based on the same sol as in
Fig.\ref{fig_tes_tw}, kriged at high resolution. 

\begin{center}--- Figure~\ref{fig_krigmap} ---\end{center}

The series of daily maps is divided into annual dust scenarios, with 669 sols each, instead of using the Martian calendar described in \ref{calendar}. The reason for this choice is that atmospheric models that use dust scenarios are often based on an integer number of sols in the year. For years with 668 sols in our sol-based Martian calendar, we add the first sol of the following year. The 669th sol of MY~24, 27, 29, therefore is the same as the first sol of, respectively, MY~25, 28, and 30. Since there are no maps available for the first part of MY~24 (before sol 225), we use the first 224 sols of MY~25 to fill the gap, therefore creating a ``hybrid'' MY~24 year. The junction at sol 225 is smoothed using a running average for the three sols before and the three sols after. It is worth mentioning that if one wants to use a scenario for a specific year in a cyclic model simulation, the first few sols and the last few sols should be averaged (e.g. with a running average) to avoid the discontinuity. 

Similarly to what we do for the irregularly gridded maps, we provide an atlas of the daily dust scenario maps as supplementary material of this paper. The structure of the atlas is the same: 669 pages (one per sol), each one having eight maps for the respective years. The dust scenarios (currently v1.5) can be downloaded from the same URL in NetCDF format. Fig.~\ref{fig_zonalmean_krig} shows zonal means built from the dust scenario maps as a function of solar longitude and latitude, as in Fig.~\ref{fig_zonalmean_grid} for the irregularly gridded maps. 

\begin{center}--- Figure~\ref{fig_zonalmean_krig} ---\end{center}

\section{Summary and future developments}
%%%%%%%%%%%%%%%%%%%%%%%%%%%%%%%%%%%%%%%%%%%%%
\label{summary}

In this paper we have described the procedures to obtain daily gridded
maps of column dust optical depth for eight Martian years,
from $\mathrm{L_{s}}\sim105^{\circ}$ in MY~24 to the end of MY~31. We have used retrieved and estimated CDODs
from three different instruments (TES, THEMIS, and MCS) aboard different
polar orbiting satellites (MGS, ODY, and MRO). Data have been collected for about
14 Earth years of spacecraft missions, from 1999 to 2013.

Observed CDODs have been firstly gridded using
the IWB methodology, which produces daily maps on a regularly distributed grid, but with missing values at some grid points. Subsequently, incomplete maps have been
interpolated onto a regular, complete grid, using the kriging method after the application of a series of procedures to replace most of the missing values with climatological values. The two datasets of IR absorption CDOD at the reference pressure of 610 Pa (irregularly gridded maps and regularly gridded ones),  
have been separated in Martian years, and made publicly available in the NetCDF format on the MCD project website hosted by the LMD (URL: \texttt{http://www-mars.lmd.jussieu.fr/}). Two atlases of the current version 1.5 of the maps are also available as supplementary material of this paper. 

Key achievements of this work are the production and analysis of a continuous, multiannual climatology of dust based on several heterogeneous datasets, as well as the assessement of uncertainties (both in the CDOD retrievals and in the resulting gridded maps), and biases. Eight Martian years of temporal and spatial (2D) dust distribution have been presented in term of interannual and intraseasonal variability, down to the daily evolution of single dust storms. We have confirmed that, from the statistical point of view, the years without global-scale storms are characterized by four phases in the solar longitude-latitude dust distribution, clearly highlighted in Fig.~\ref{fig_dodtypyear}. This figure represents a `typical' climatological year for the dust optical depth, on top of which the annual optical depth is locally increased by the evolution of single dust storms, particularly in the second half of the year.  

Future developments of this work include the update of the climatology with the addition of new observations, not only for Martian years beyond MY~31, but also for those already described in this paper. The MCS team expects to carry out retrievals of CDOD (as additional standard product) using nadir and off-nadir observations, which would greatly improve the quality of the dataset and reduce the biases identified in this work. The use of estimated CDODs from fitting top-of-atmosphere brightness temperatures with a GCM can be another important source of information to be used in the gridding procedure, in order to increase the number of valid grid points and reduce the need for the kriging interpolation in some areas. Reliable retrievals of CDOD from the Planetary Fourier Spectrometer (PFS) observations aboard Mars Express would be invaluable to improve the quality of the maps in MY~27 and MY~28, when only sparse THEMIS observations are available. The IWB procedure would be particularly adapted to grid PFS observations, which are fairly heterogeneous in local time.    

The use of data assimilation in a dynamical model of the dust cycle is the obvious improvement of the simple spatial kriging technique described in this work to produce complete dust scenarios, with ensemble methods providing the capability to estimate uncertainties. Despite possible use of data assimilation techniques as `optimal interpolators' of dust observations (or other types of aerosols and variables, e.g. \citet{ste14}), gridding retrievals from satellite orbits with IWB procedure can retain its key advantages of 1) being easy to implement, 2) not requiring a model, 3) providing statistical estimates of uncertainties based on those of the retrievals, and 4) giving immediate access to statistical analysis of the spatio-temporal evolution of events. Furthermore, this gridding technique could be applied to fields beyond the study of Martian dust climatology, where data assimilation techniques might not be accessible or might be more difficult to implement.

\section*{Acknowledgments}
\label{ack}
LM is indebted to many people whose direct or indirect help made it possible to
write this paper. A particular mention is due to Aymeric Spiga, who always
believed this paper would be finished one day, and provided time and competence
during long discussion sessions.  
LM is grateful to Mathieu Vincendon, Tanguy Bertrand, Frank Daerden for
feedback on earlier versions of the dust scenario maps, and to Helen Wang for initially guiding us through the MARCI images. 
%We are also particularly grateful to Mathieu Vincendon and another anonymous reviewer for their comments and suggestions, which helped to improve the paper. Finally, LM wishes to thank his wife, H\'elo\"ise, for accepting the loss of many days of vacation during the period he worked on this paper...and yet remaining his wife! For a paper where all authors are men, there was no better day to be initially submitted than the 2014 International Women's Day (IWD)! If the reader wants to know which sol it was on Mars, then \ref{calendar} provides the answer.

\appendix

\section{A sol-based Martian calendar}
\label{calendar}
A Martian year has 668.5921 mean solar days (sols) or, with good approximation, 668.6 sols. The solar longitude resets to zero by definition at the beginning of each new year (spring equinox).

The scientific community primarily have been using a Martian calendar based on the combination of Martian year and solar longitude, since data are usually averaged over a few solar longitude degrees. This combination is not satisfactory in our work, because our series of maps uses the sol as time variable. If we want to devise a Martian calendar based on sols, we need to keep track of the fraction of a sol left at the end of each Martian year, similarly to what is done for the Earth calendar. Many sol-based calendars have been proposed for Mars, with different solutions for the structure of months and weeks, and for the way they deal with the leap years. We mention here only the example of the Darian calendar \citep[first described in a paper published by ][]{gan86}, which has been proposed to serve the needs of possible future human settlers on Mars.

For scientific purposes, and in the interest of simple time-keeping, we have devised for the present work our own sol-based Martian calendar. Since there are approximately 668.6 sols in a Martian year, every 5 years the number of elapsed sols is very close to the integer 3343. It is convenient, therefore, to define cycles of 5 Martian years, with a number of sols respectively of 669, 668, 669, 668, and 669 (a total of 3343 sols). By doing this, the end of each cycle is (without approximation) only 0.04 sol shorter than an integer number of sols, which accumulates to 1/4 of a sol after 30 Martian years (a reasonable amount for scientific purposes). Furthermore, assuming the first sol of a year starts when it is midnight at the prime meridian (00:00 MUT), if there is a time bias $t_{b}$ between the first sol and $\mathrm{L_{s}}=0^{\circ}$ at the beginning of a 5-year cycle, there is again the same bias at the beginning of the following cycle, if we approximate the Martian year with 668.6 sols. Thanks to this 
property, the time biases between our sol-based calendar and a solar longitude-based calendar follow a repeatable pattern, and they are never too large for a reasonable number of Martian years, if $t_{b}$ is properly set at the beginning of MY~1. 

\citet{cla00} proposed an arbitrary convention to numerate the Martian years starting from 11th April 1955 (MY~1). This date corresponds to the Martian northern spring equinox preceding the planet-encircling dust storm of 1956. This storm marks the beginning of the new period of scientific observation and exploration of the planet Mars. Using this convention, all scientific data can be easily compared using a calendar based on positive Martian years. Since the introduction of this convention, most of the scientific papers on Mars adopted the numeration introduced by Clancy et al., according to which $\mathrm{L_{s}}=0^{\circ}$ in MY~1 occurred at 08:31 UTC on 11th April 1955. For a solar longitude-based calendar, every new Martian year since this date is marked by the occurrence of the northern hemisphere spring equinox. 

For the sol-based calendar introduced in the present paper, we maintain the numeration of the Martian year introduced by \citet{cla00}, but we have to start our MY~1 at a slightly different time, because at $\mathrm{L_{s}}=0^{\circ}$ in MY~1 the MUT time (local mean solar time) was 13:26 rather than 00:00. The first sol of MY~1 in our calendar starts on 11th April 1955 at 19:22 UTC, when it was $\mathrm{L_{s}}=0.2^{\circ}$ and 00:00 MUT on Mars, i.e. $t_{b}=0.44$ sol. By the beginning of MY~26, this bias has reduced to $t_{b}=0.24$, therefore the sol-based calendar we use is less than 6 hours late with respect to the calendar based on solar longitude at the spring equinox of MY~26, and less than 5 hours late at the beginning of MY~31 (i.e the beginning of the following 5-year cycle). This bias is satisfactory for the climatology from MY~24 to MY~31, and we make the approximation that $t_{b}=0.24$~sol ($0.12^{\circ}$ solar longitude) for the three cycles we deal with in this paper. By doing so, we can easily 
subtract a constant 
offset at the new year's sol in MY~21, 26 and 31, to find the solar longitude of the corresponding spring equinox, with an error of about one hour in MY~21 and 31 with respect to MY~26. In Table~\ref{tab2}, we summarize the number of sols for each Martian year and the solar longitude corresponding to each new year's sol. 

\begin{center}--- Table~\ref{tab2} ---\end{center}

%At the date and time of submission of this paper (8th March 2014, about 12 UTC), IWD on Earth, we were entering sol 215 of MY~32 on Mars,  $\mathrm{L_{s}}=99.5^{\circ}$ (northern hemisphere summer)\footnote{Calculated using the NASA-GISS Mars24 applet, \texttt{http://www.giss.nasa.gov/tools/mars24/}, and the MCD Earth-to-Mars date applet, \texttt{http://www-mars.lmd.jussieu.fr/mars/time/martian\_time.html}}.
%At the date this paper was finished (29th February 2014), it was $\mathrm{L_{s}}~96^{\circ}$ in MY~32\footnote{According to the NASA-GISS Mars24 applet v7.0.1a, \texttt{http://www.giss.nasa.gov/tools/mars24/}, although the MCD Earth-to-Mars date applet (\texttt{http://www-mars.lmd.jussieu.fr/}) disagrees.}.



\pagebreak

\pagebreak
\section*{Tables}

\begin{sidewaystable}
\centering
\begin{tabular}{|c||c||c|c|c|c|c|c|c|}
\hline
           & Spatial grid            & TW              & $d_{cutoff}$      & $S_{min}$          & $S_{max}$             & $d_{thr}$            & $N_{thr}$       & $(\sigma_{\tau}/\tau)_{thr}$\\       
           & (lon-lat)               & (sols)          & ($10^2$~km)       & ($10^2$~km)        & ($10^2$~km)           & ($10^2$~km)          &                 &   \\      
\hline \hline
TES        & $6^\circ\times3^\circ$  & 1, 3, 5, 7      & 5, $8^{2,3,4}$    & $1.5^{1,2,3,4}$    & 1.5, $3^{2,3,4}$      & 2, $3^{2,3,4}$       & $3^{1,2,3,4}$   & $0.4^{1,2,3,4}$\\  
TES+THEMIS &                         &                 &                   &                    &                       &                      &                 &  \\  \hline
THEMIS     & $6^\circ\times5^\circ$  & $3^{1,2}$, 5, 7 & $12^{1,2,3,4}$    & $1.5^{1,2,3,4}$    & $3^{1,2,3,4}$         & 4, 10, 15, 10        & $2^{1,2}$, $3^{3,4}$  & $0.4^{1,2,3,4}$\\ \hline
MCS+THEMIS & $6^\circ\times5^\circ$  & 1, 3, 5, 7      & 5, $8^{2,3,4}$    & $1.5^{1,2,3,4}$    & 1.5, $3^{2,3,4}$      & 2, $3^{2,3,4}$       & $3^{1,2,3,4}$   & $0.4^{1,2,3,4}$\\
\hline
\end{tabular}
\caption{Parameters for the gridding procedure.This table summarizes the parameters used in the gridding
procedure with IWB for the four combinations of datasets and for the different
iterations. We have used 4 iterations of the procedure, mainly changing the time
window. The superscripts indicate that the same parameter has been used for more
than one iteration (`1' indicates the first iteration, etc.). Other parameters
not included in this table are kept fixed in this work, i.e. the time resolution
of the maps is 1 sol, the value of the $R$ function at the extrema of the TW is
0.05.}
\label{tab1}
\end{sidewaystable}

\pagebreak
\begin{table}
\centering
\begin{tabular}{|c|c|c|c|}
\hline 
5-year cycle & Martian year & Number of sols & New year's   \\  
number       &              &                & solar longitude  \\  \hline \hline
5 & 24 & 668 & 359.98 \\  \hline
5 & 25 & 669 & 359.67 \\  \hline
6 & 26 & 669 & 359.88 \\  \hline
6 & 27 & 668 & 0.08   \\  \hline
6 & 28 & 669 & 359.78 \\  \hline
6 & 29 & 668 & 359.98 \\  \hline
6 & 30 & 669 & 359.67 \\  \hline
7 & 31 & 669 & 359.88 \\  \hline
\end{tabular}
\caption{The sol-based Martian calendar used in the present work. MY~1 new year's sol (00:00 MUT) is on 11th April 1955 at 19:22 UTC.}
\label{tab2}
\end{table}

\pagebreak
\section*{Figure captions}

\begin{figure}[H]
\centering
  \includegraphics[width=\textwidth]{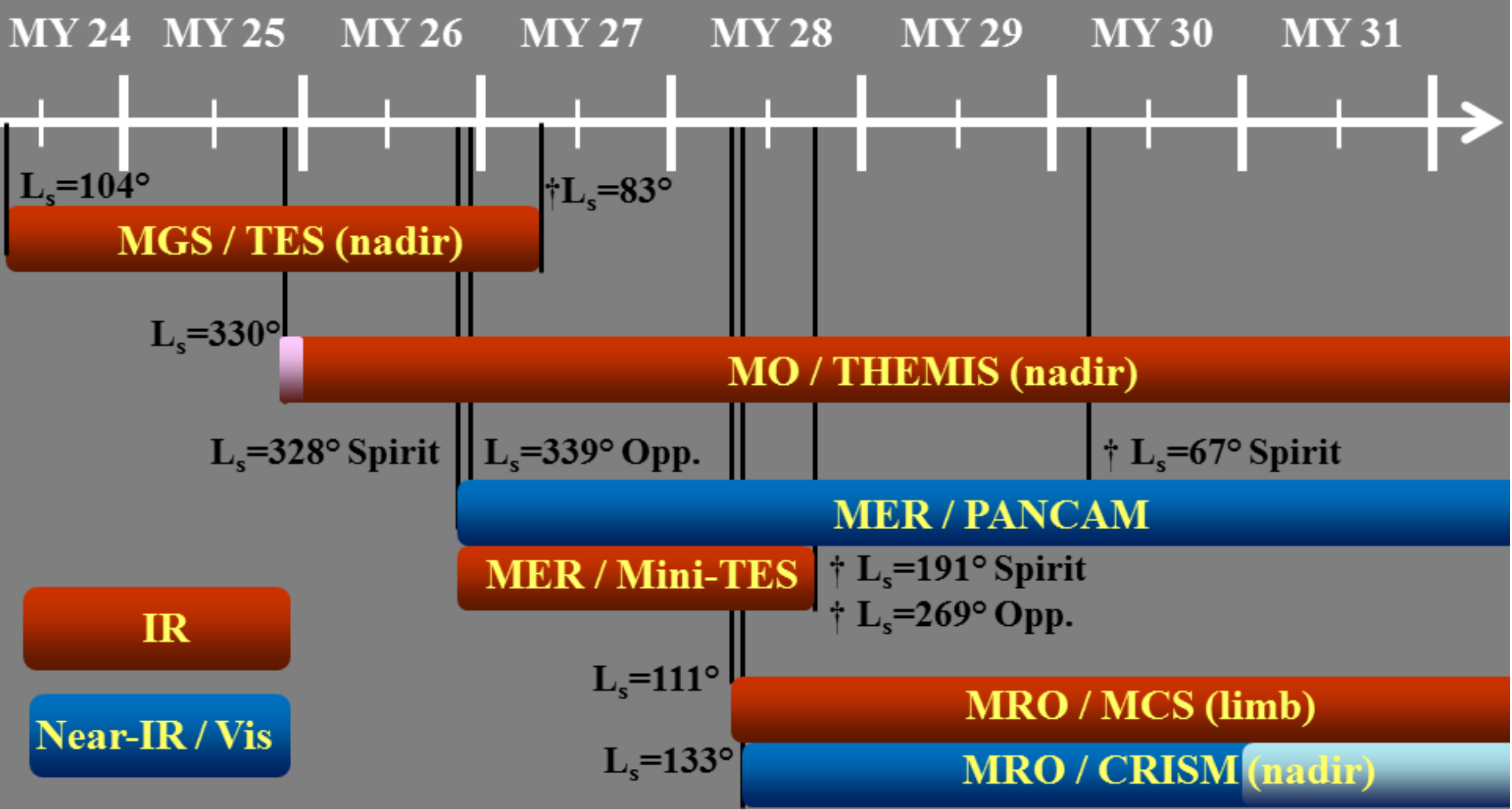}
  \caption{}\label{fig_timeline}
\end{figure}

\noindent\textbf{Figure~\ref{fig_timeline}} \,\,A summary of the time coverage of all instruments used in this work, and the time limits within which observations are available. Lighter colors in the bars indicate the periods when observations from that particular instrument are available but either there are no CDOD retrievals available or we do not currently use them.

\begin{figure}[H]
\centering
 \includegraphics[width=\textwidth]{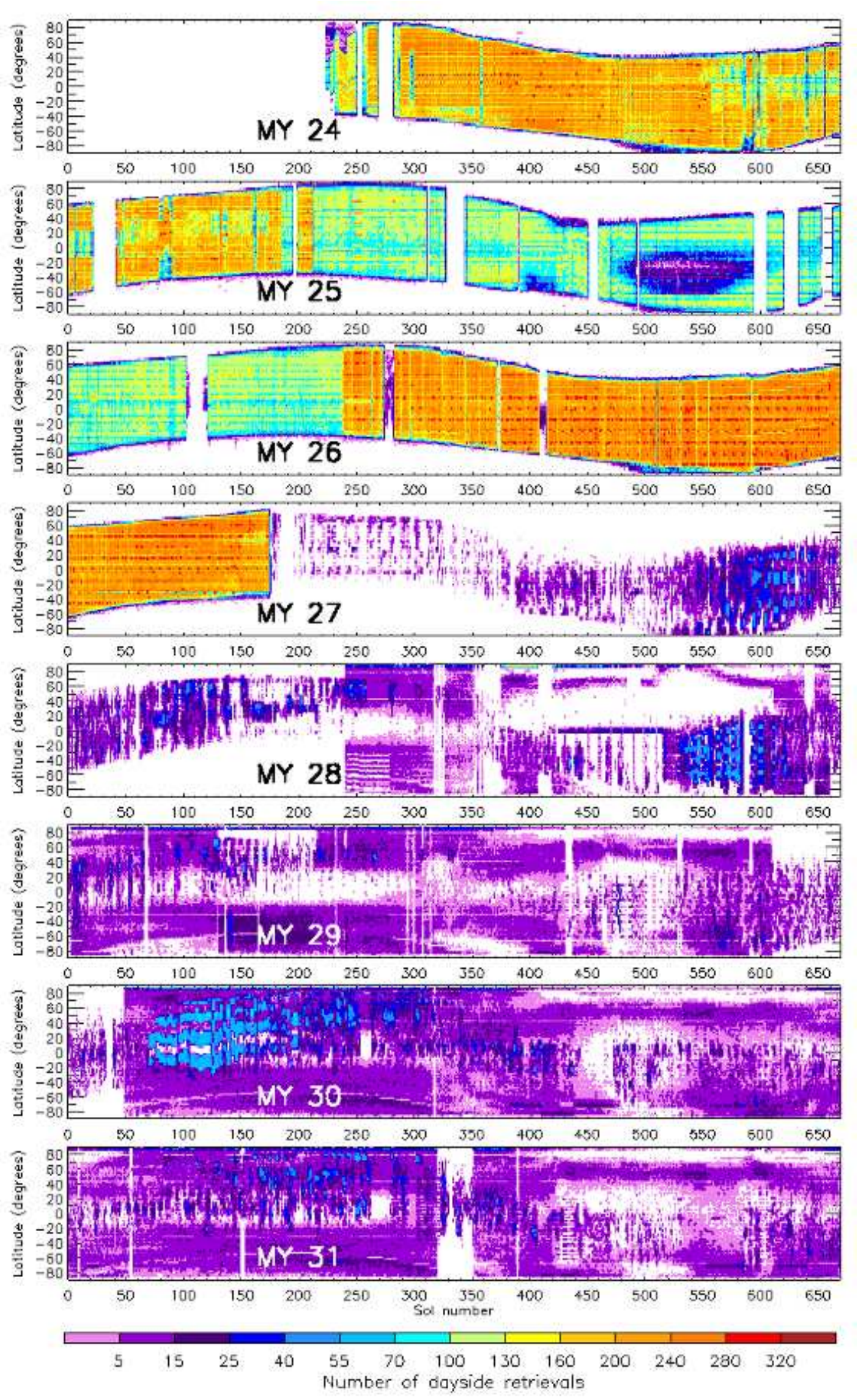}
 \caption{}\label{fig_retrievalnumberday}
\end{figure}

\noindent\textbf{Figure~\ref{fig_retrievalnumberday}} \,\,The number of dayside retrievals of column dust optical depth in absorption at 9.3~$\mu\mathrm{m}$ (local times between 06:00 and 18:00) passing the quality control procedure described in the text. The number of retrievals is summed in 1 sol~$\times$~$2^{\circ}$~latitude bins, and plotted for each year as a function of sol and latitude. TES observed at higher spectral resolution (5 cm$^{-1}$) between sol $\sim210$ in MY~25 and sol $\sim240$ in MY~26, therefore there are considerably less observations in this period, because the high spectral resolution data take twice as long to acquire as the low spectral resolution. Note that THEMIS retrievals are summed together with TES retrievals and MCS retrievals, when available.

\begin{figure}[H]
\centering
\includegraphics[width=\textwidth]{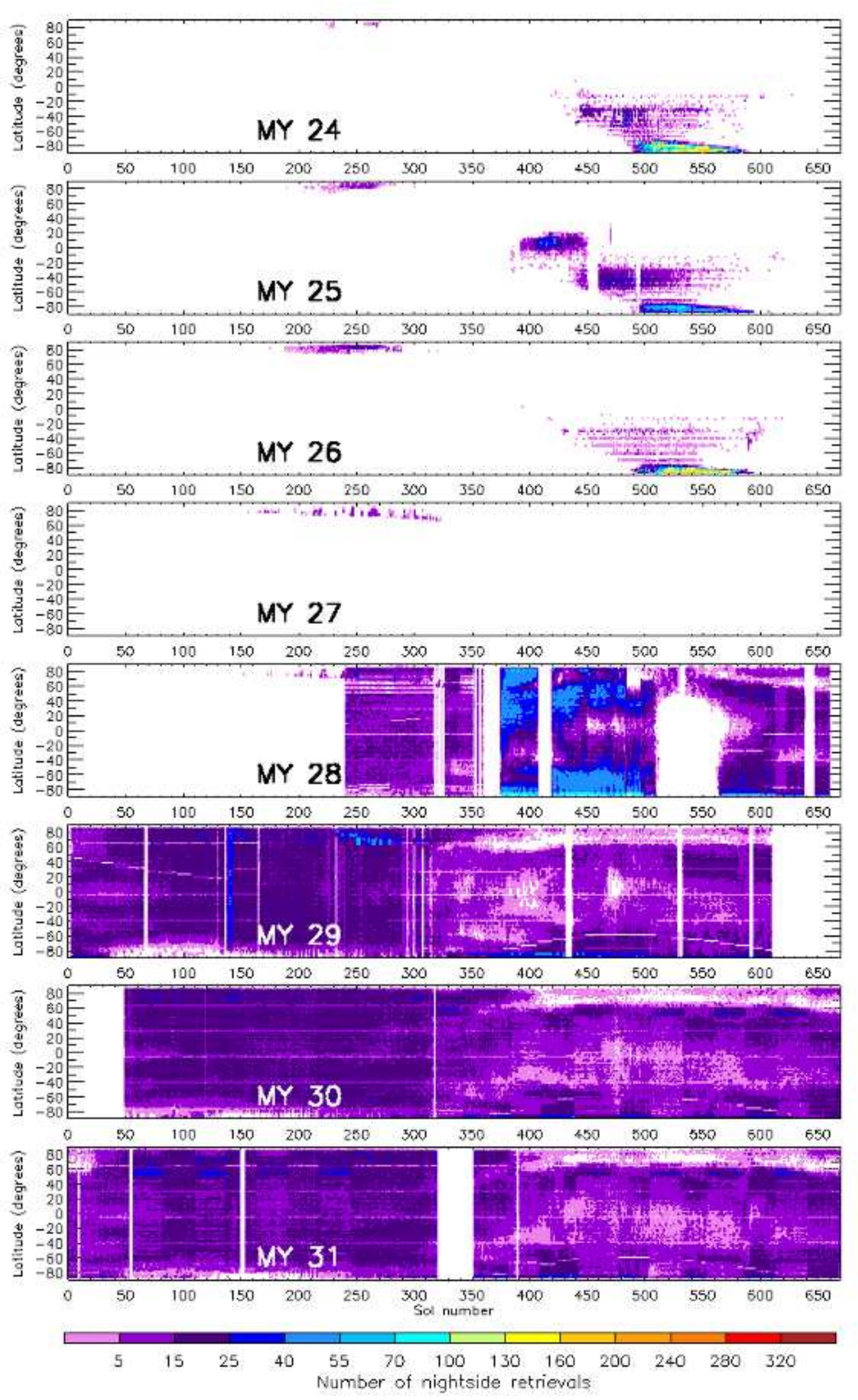}
 \caption{}\label{fig_retrievalnumbernight}
\end{figure}

\noindent\textbf{Figure~\ref{fig_retrievalnumbernight}} \,\,Same as in Fig.~\ref{fig_retrievalnumberday}, but for nightside retrievals, with local times between 00:00 and 06:00, and between 18:00 and 24:00.

\begin{figure}[H]
\centering
  \noindent\includegraphics[width=0.65\textwidth]{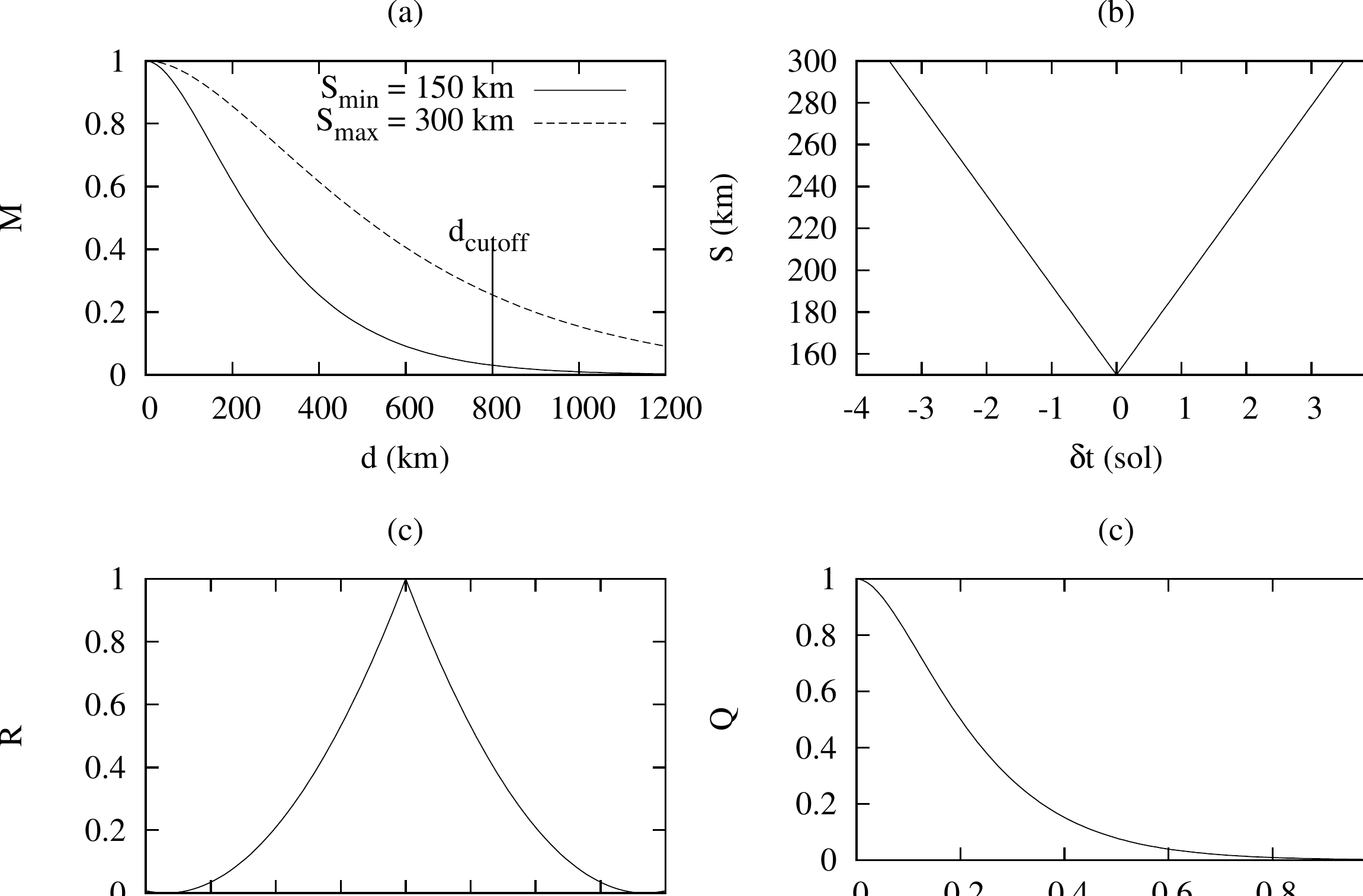}
  \caption{}\label{fig_weights}
\end{figure}

\noindent\textbf{Figure~\ref{fig_weights}} \,\,Plots of the weighting functions used in Eq.~\ref{eq2}. Panel (a) shows the distance weight $M$ as a function of $d$ (see Eq.~\ref{eq3}) for two
different values of correlation scale $S$, corresponding to $S_{min}=150$~km and
$S_{max}=300$~km. The cut-off distance $d_{cutoff}$ is drawn here at 800 km as
an example. The $S$ function is plotted in panel (b), as a function of $\delta
t$, defined over the range $-3.5<\delta t<3.5$ for TW=7 sol. In panel (c) we
plot the time weight function $R$ (see Eq.~\ref{eq3tris}) for the same TW=7 sol.
The value of the $R$ function at the extrema of the time window interval is
fixed to 0.05 in this work. Finally, panel (d) shows the plot of the uncertainty
weight $Q$ (see Eq.~\ref{eq4}) as a function of the relative uncertainty
$\sigma_{tau}/\tau$, where $\sigma$ is the standard deviation of a single column dust optical depth retrieval $\tau$.

\begin{figure}[H]
\centering
  \noindent\includegraphics[width=\textwidth]{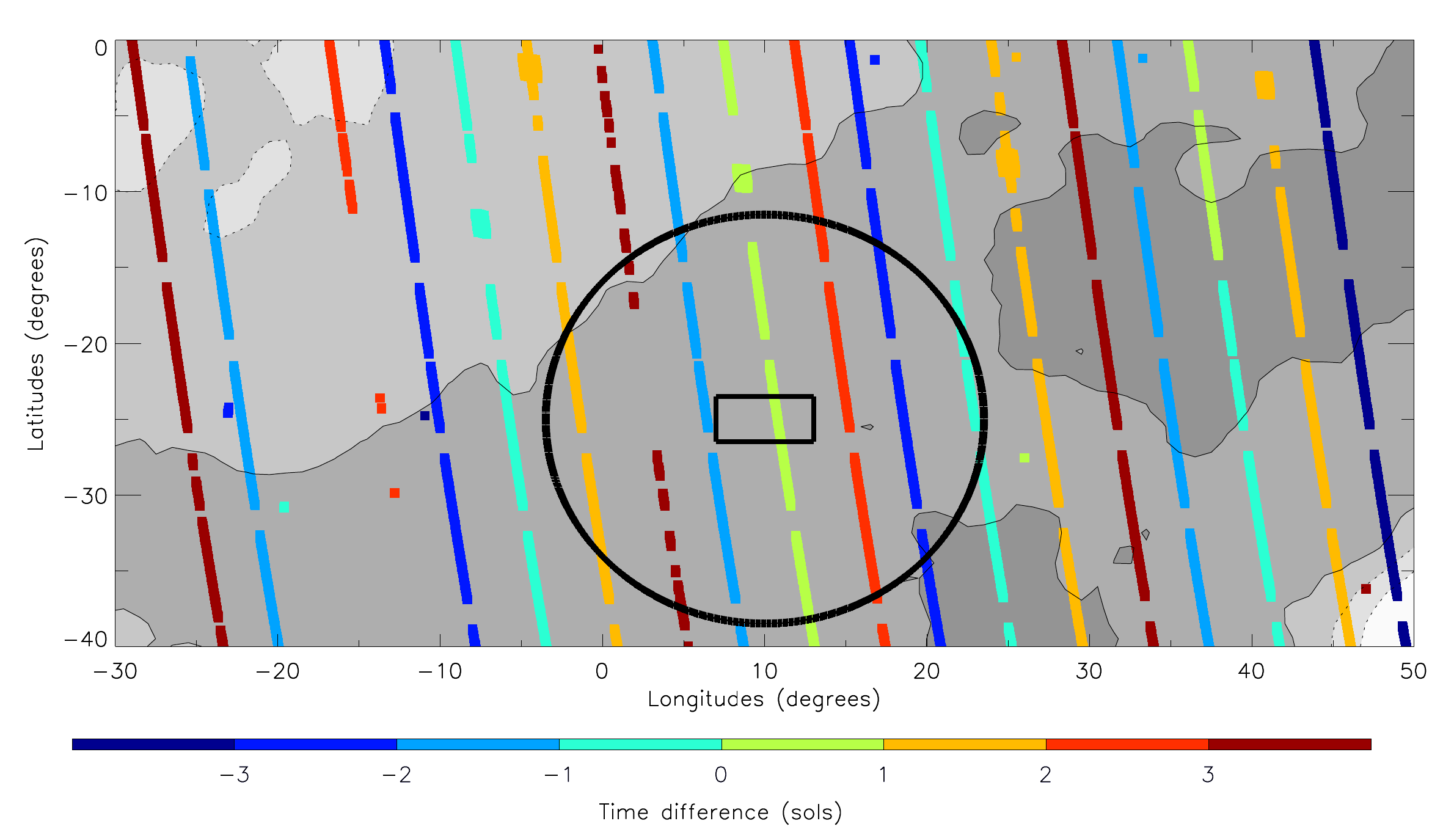}
  \caption{}\label{fig_tes_time}
\end{figure}

\noindent\textbf{Figure~\ref{fig_tes_time}} \,\,The difference between the time of each TES
observation and noon at $0^{\circ}$ longitude in sol 449, $\mathrm{L_{s}}=227.8^{\circ}$, MY~24 (the `current time'). The observations are shown within a time window of 7 sols centered around the current time. The picture is
a zoomed view centered around $10^{\circ}$ longitude and $-25^{\circ}$ latitude.
The black square shows a $6^{\circ}\times3^{\circ}$ longitude-latitude box whereas
the black circle indicates the 800 km distance range ($d_{cutoff}$) we use in
the weighted binning for the TES dataset when the time window is 7 sols (see
Table~\ref{tab1}). We apply the time and spatial weighting within this circle, as detailed in Section~\ref{bingrid}.

\begin{figure}[H]
\centering
  \noindent\includegraphics[width=\textwidth]{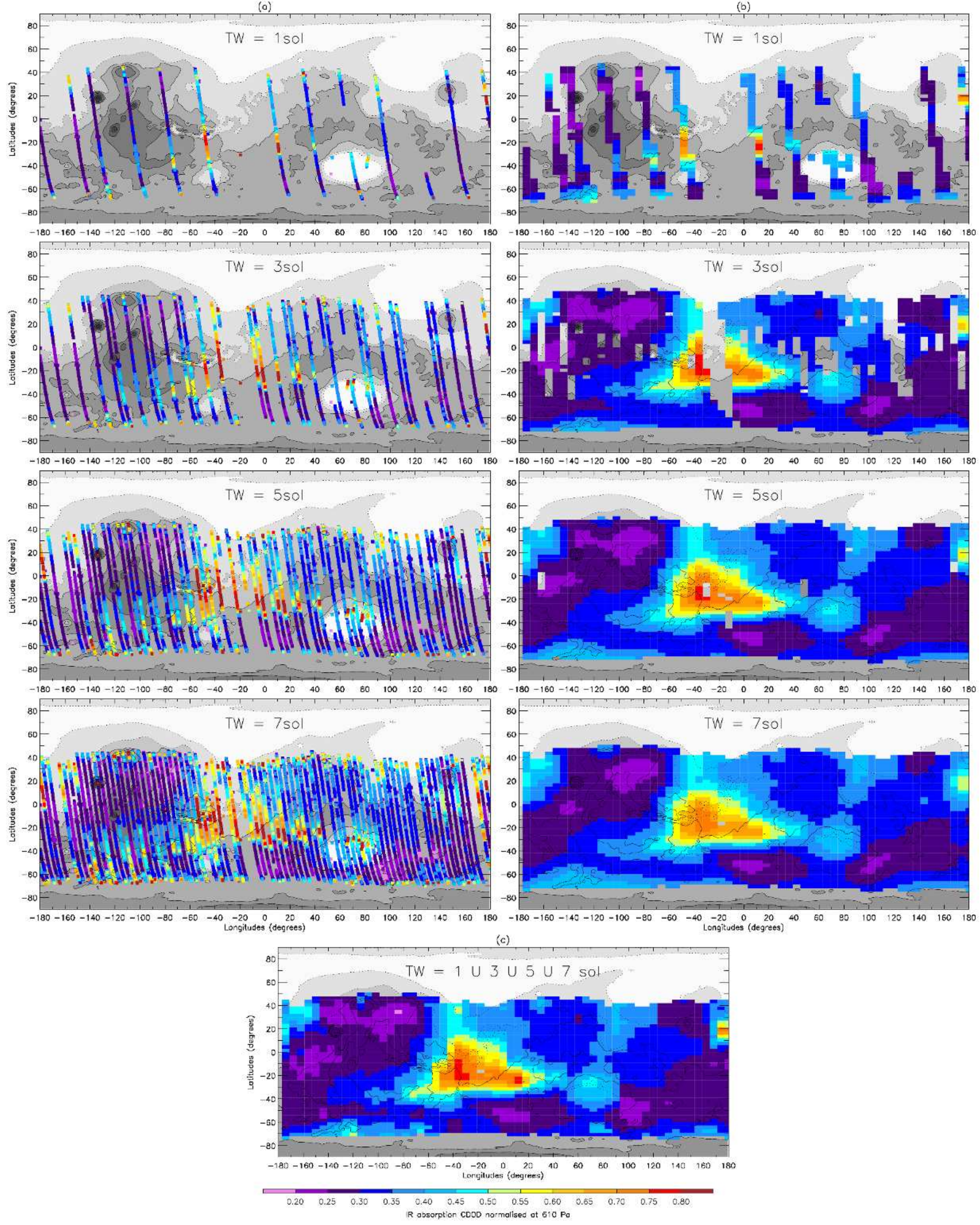}
  \caption{}\label{fig_tes_tw}
\end{figure}

\noindent\textbf{Figure~\ref{fig_tes_tw}} \,\,(a) The panels from top to bottom show TES
observations accumulated within time windows of 1, 3, 5 and 7 sol, centred
around noon at $0^{\circ}$ longitude in sol 449. (b) The panels
from top to bottom show the corresponding partial results of the
application of the weighted binning for each specific time window. Finally, in panel (c) we show the result of the iterative application of all 4 TW, where the valid grid
points in each iteration are not overwritten. There are 60x60 grid
points in the gridded maps, separated by $6^{\circ}\times3^{\circ}$ in longitude
and latitude. Missing values are assigned to bins where the acceptance criterion
is not satisfied (see Table~\ref{tab1}).

\begin{figure}[H]
\centering
  \noindent\includegraphics[width=\textwidth]{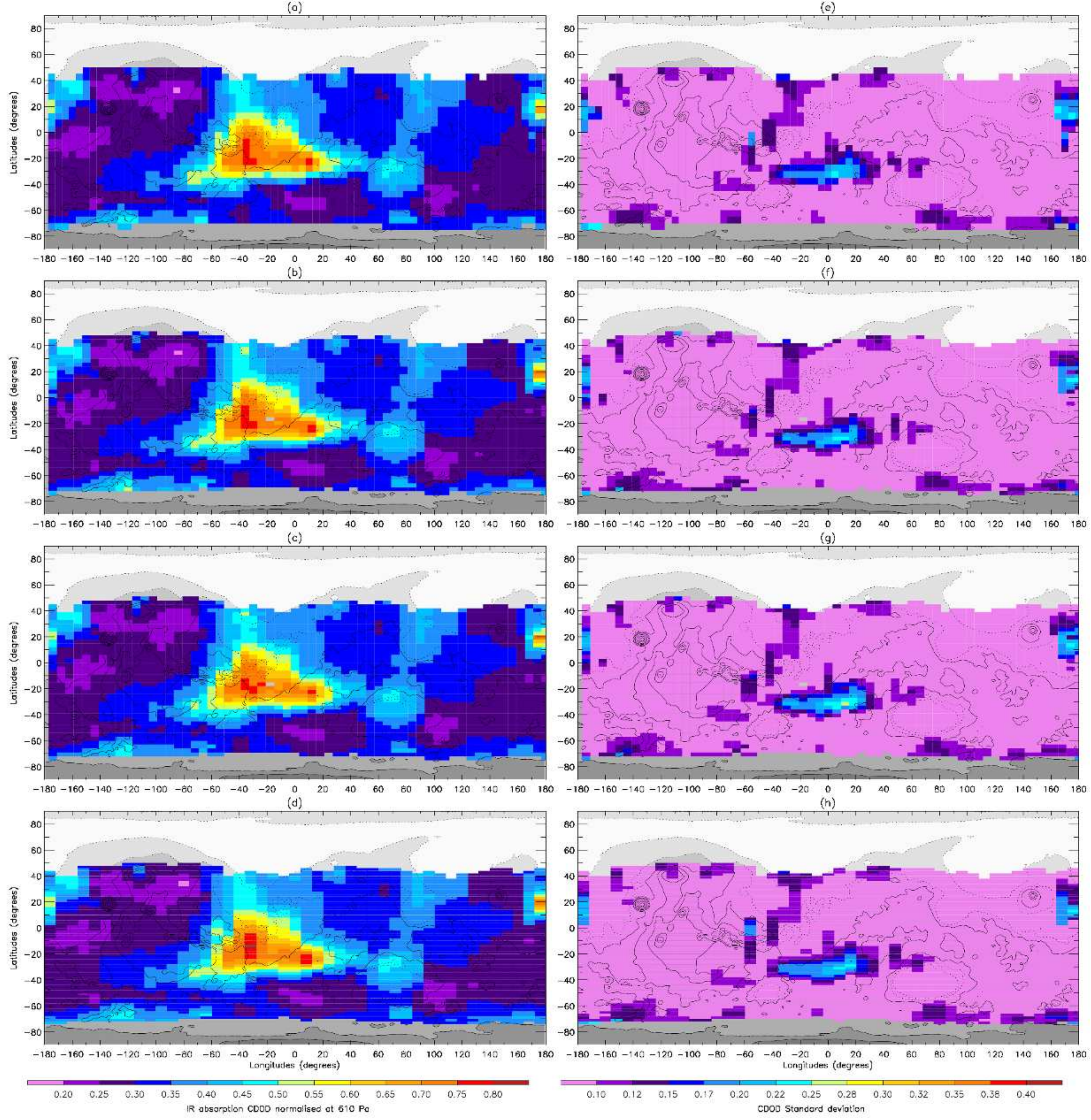}
  \caption{}\label{fig_my24_restests}
\end{figure}

\noindent\textbf{Figure~\ref{fig_my24_restests}} \,\,Tests on the sensitivity of the gridding procedure on the spatial resolution. We use the same sol as in Fig.~\ref{fig_tes_tw}. Panel (a) and (f) show the case with $5^{\circ}\times5^{\circ}$ longitude-latitude resolution, respectively in the average and in the standard deviation fields; panel (b) and (g) are for the case with $6^{\circ}\times3^{\circ}$ resolution; panel (c) and (h) are for the case with $6^{\circ}\times3^{\circ}$ resolution but longer range for the space weighting ($S_{min}=250$~km and $S_{max}=400$~km); panel (d) and (i) are for the case with $8^{\circ}\times2^{\circ}$ resolution.

\begin{figure}[H]
\centering
  \noindent\includegraphics[width=\textwidth]{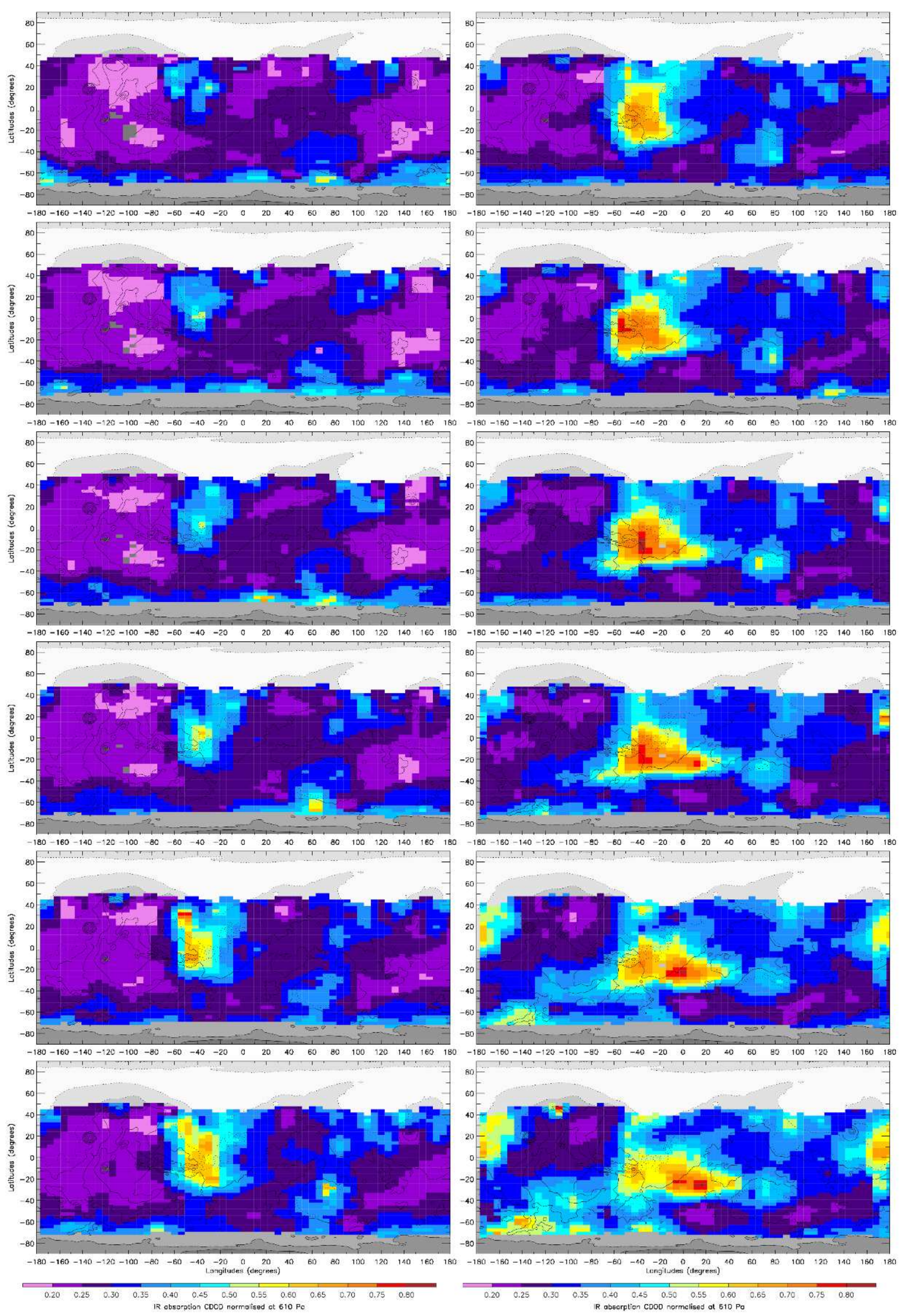}
  \caption{}\label{fig_my24flush_evol}
\end{figure}

\noindent\textbf{Figure~\ref{fig_my24flush_evol}} \,\,Evolution of the MY~24, $\mathrm{L_{s}}\sim227^{\circ}$ flushing storm. Each panel shows gridded column dust optical depth (in absorption at 9.3~$\mu\mathrm{m}$) at the reference pressure level of 610 Pa. From top left to bottom right, maps are provided from sol 440 to sol 451 (i.e. from $\mathrm{L_{s}}\sim221.4^{\circ}$ to $\mathrm{L_{s}}\sim228.4^{\circ}$). In all panels, MUT is noon and solar longitudes are provided accordingly. See also \ref{calendar} for the description of the sol-based Martian calendar we use in this paper.

\begin{figure}[H]
\centering
  \noindent\includegraphics[width=\textwidth]{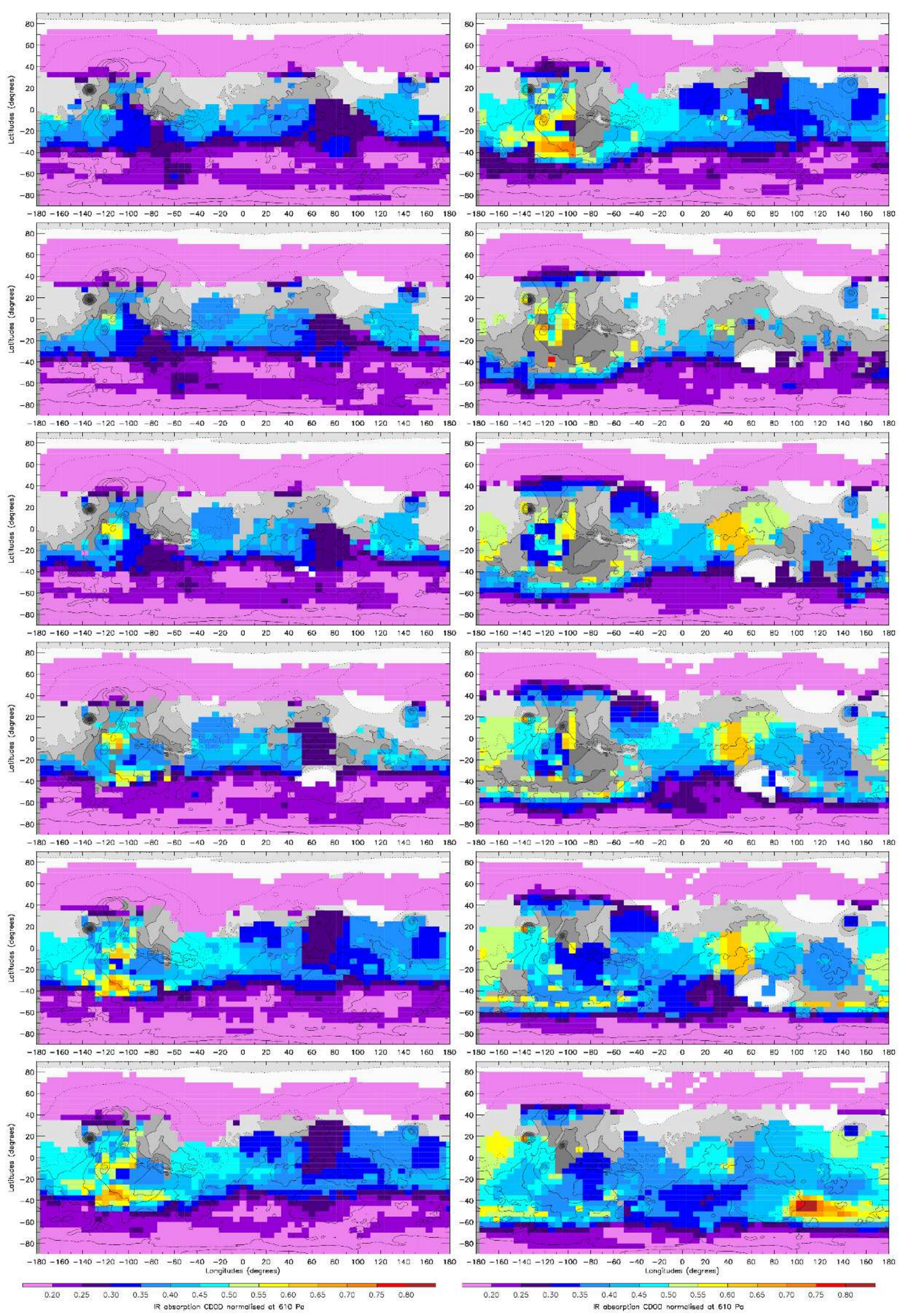}
  \caption{}\label{fig_my29flush_evol}
\end{figure}

\noindent\textbf{Figure~\ref{fig_my29flush_evol}} \,\,Evolution of the MY 29, $\mathrm{L_{s}}\sim235^{\circ}$ regional storm. As in Fig.~\ref{fig_my24flush_evol}, each panel shows gridded column dust optical depth (in absorption at 9.3~$\mu\mathrm{m}$) at reference pressure level of 610 Pa. From top left to bottom right, maps are provided from sol 457 to sol 468 (i.e. from $\mathrm{L_{s}}\sim232.3^{\circ}$ to $\mathrm{L_{s}}\sim239.4^{\circ}$). 

%\noindent\textbf{Figure~\ref{fig_correlation_grid_TES}} \,\,

%\noindent\textbf{Figure~\ref{fig_correlation_grid_MCS}} \,\,

\begin{figure}[H]
\centering
  \noindent\includegraphics[width=\textwidth]{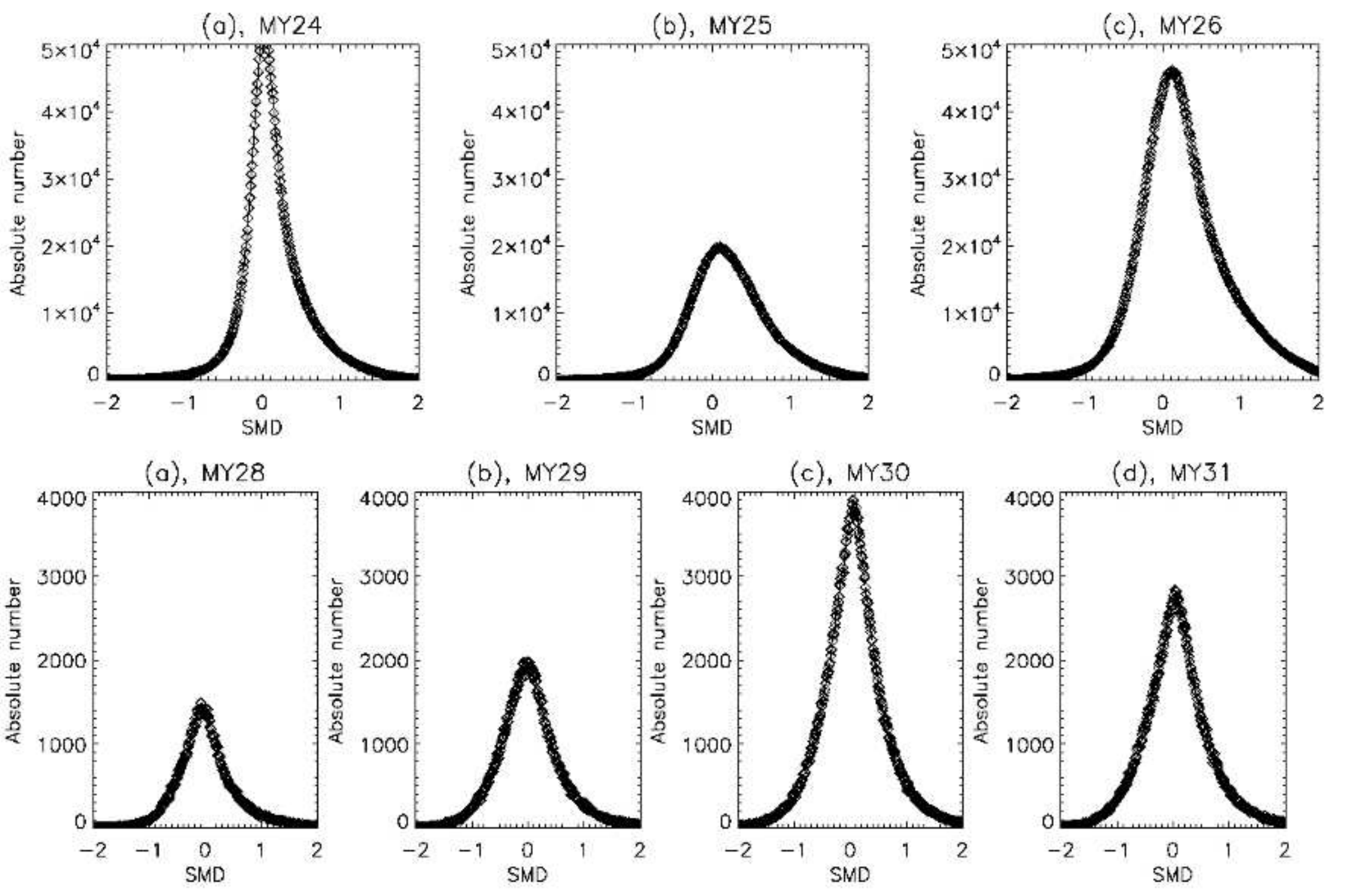}
  \caption{}\label{fig_histo_grid}
\end{figure}

\noindent\textbf{Figure~\ref{fig_histo_grid}} \,\,Histograms of SMD values (Eq.~\ref{eq5}) for TES and MCS datasets in different Martian years: (a) TES, MY~24;  (b) TES, MY~25; (c) TES, MY~26; (d) MCS, MY~28; (e) MCS, MY~29; (f) MCS, MY~30; (g) MCS, MY~31.

\begin{figure}[H]
\centering
  \noindent\includegraphics[width=\textwidth]{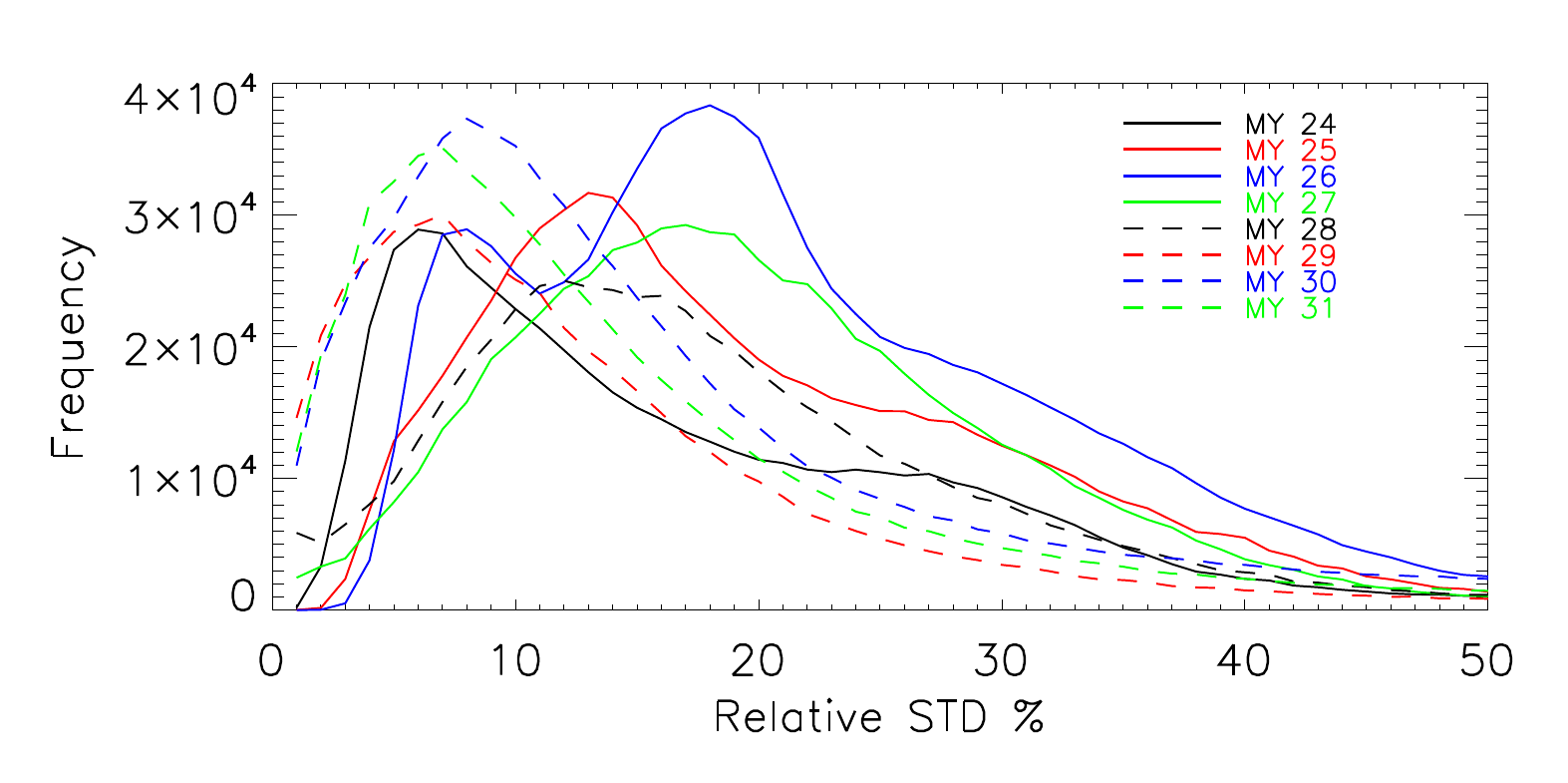}
  \caption{}\label{fig_percrelerr}
\end{figure}

\noindent\textbf{Figure~\ref{fig_percrelerr}} \,\,Plots of relative standard deviation curves for different Martian years. The relative standard deviation is defined as the ratio between the weighted standard deviation and the weighted average calculated, respectively, with Eq.~\ref{eq2} and Eq.~\ref{eq2bis}, for each valid grid point.

%\noindent\textbf{Figure~\ref{fig_correlation_grid_CRISM}} \,\,

\begin{figure}[H]
\centering
  \noindent\includegraphics[width=\textwidth]{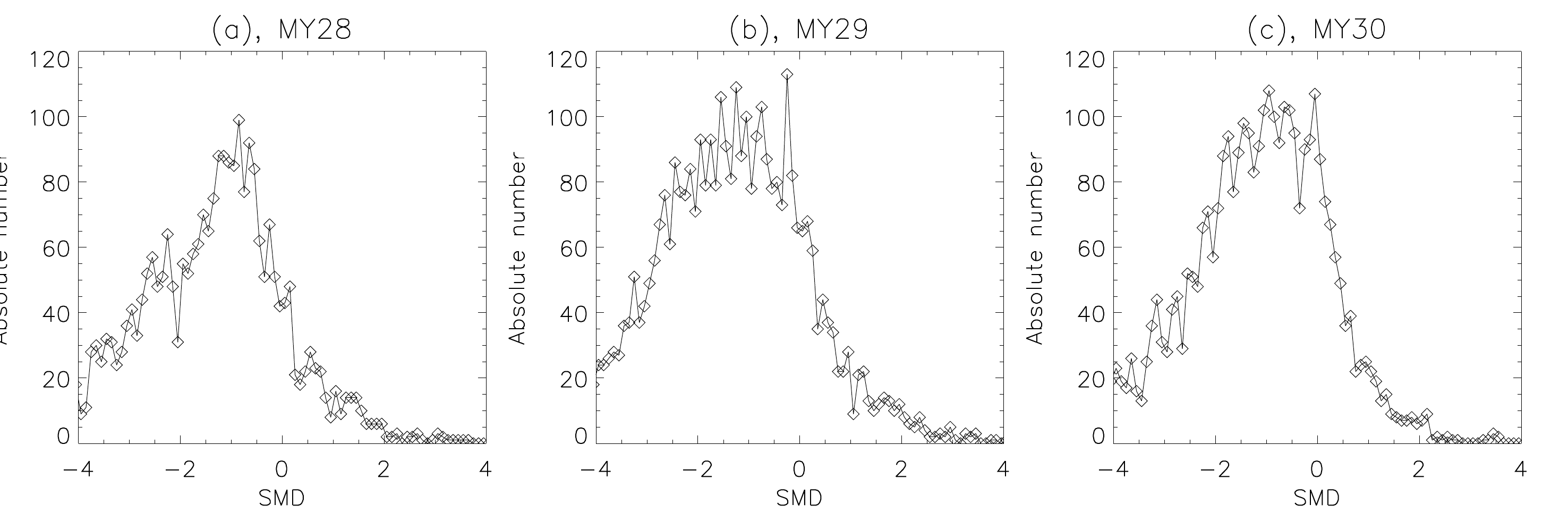}
  \caption{}\label{fig_histo_grid_CRISM}
\end{figure}

\noindent\textbf{Figure~\ref{fig_histo_grid_CRISM}} \,\,Histograms of SMD values (Eq.~\ref{eq5}) for the CRISM dataset in different Martian years: (a) MY~28;  (b) MY~29; (c) MY~30.

\begin{figure}[H]
\centering
  \noindent\includegraphics[width=\textwidth]{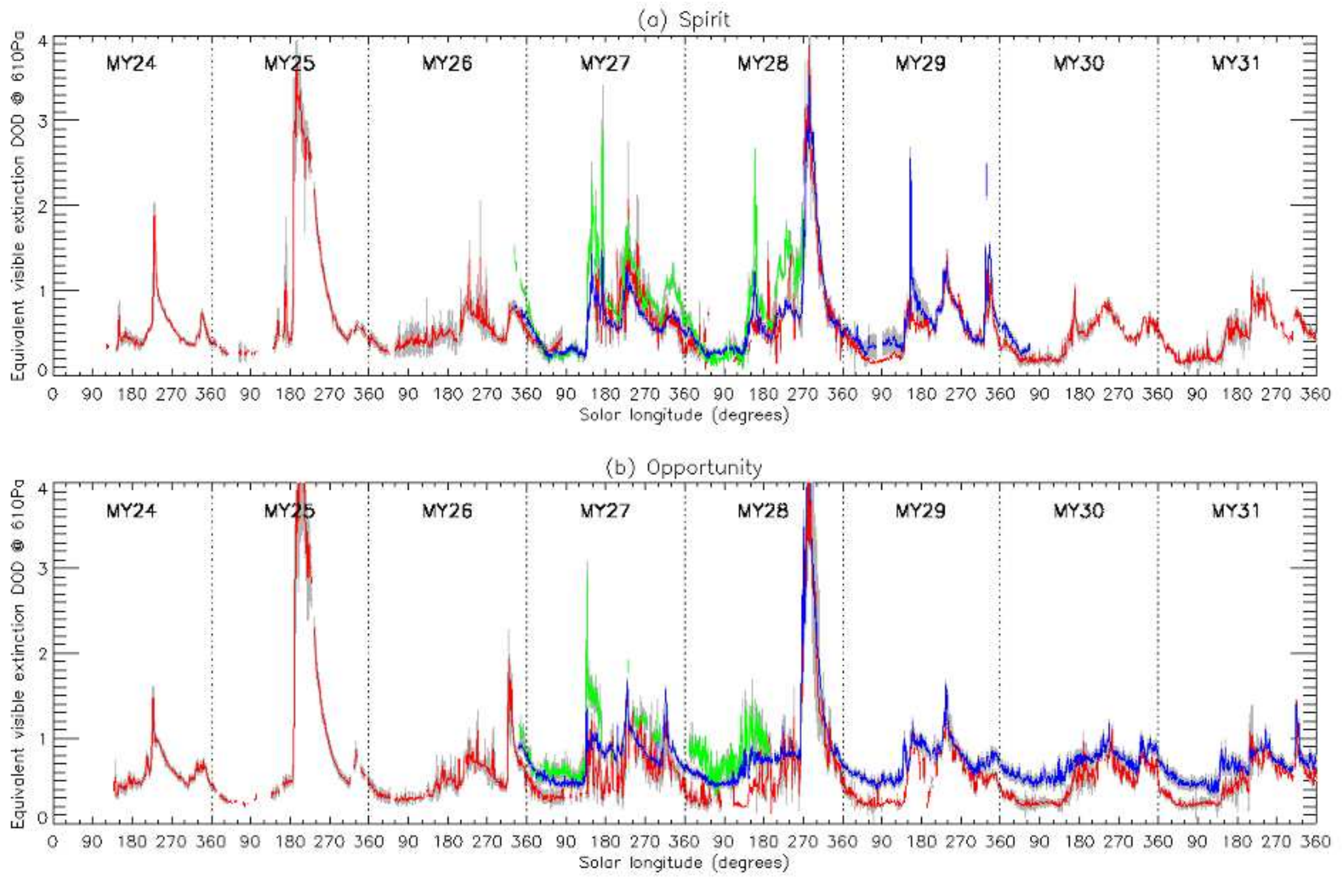}
  \caption{}\label{fig_MERtimeseries}
\end{figure}

\noindent\textbf{Figure~\ref{fig_MERtimeseries}} \,\,Plot of column dust optical depth time series at (a) Spirit site in Gusev crater, and (b) Opportunity site in Meridiani Planum. The red curves are the time series of values interpolated from the daily gridded maps obtained with IWB procedure (only values where 4 neighbor grid points are valid). The green curves are sol-averaged values from MER Mini-TES, whereas the blue curves are sol-averaged values from MER PanCam. Corresponding uncertainty values are drawn as gray envelopes. Values interpolated from the gridded maps and Mini-TES values are multiplied by 2.6 to convert them from absorption at 9.3~$\mu\mathrm{m}$ to equivalent visible, to compare to PanCam values at 800 nm.   

\begin{figure}[H]
\centering
  \noindent\includegraphics[width=\textwidth]{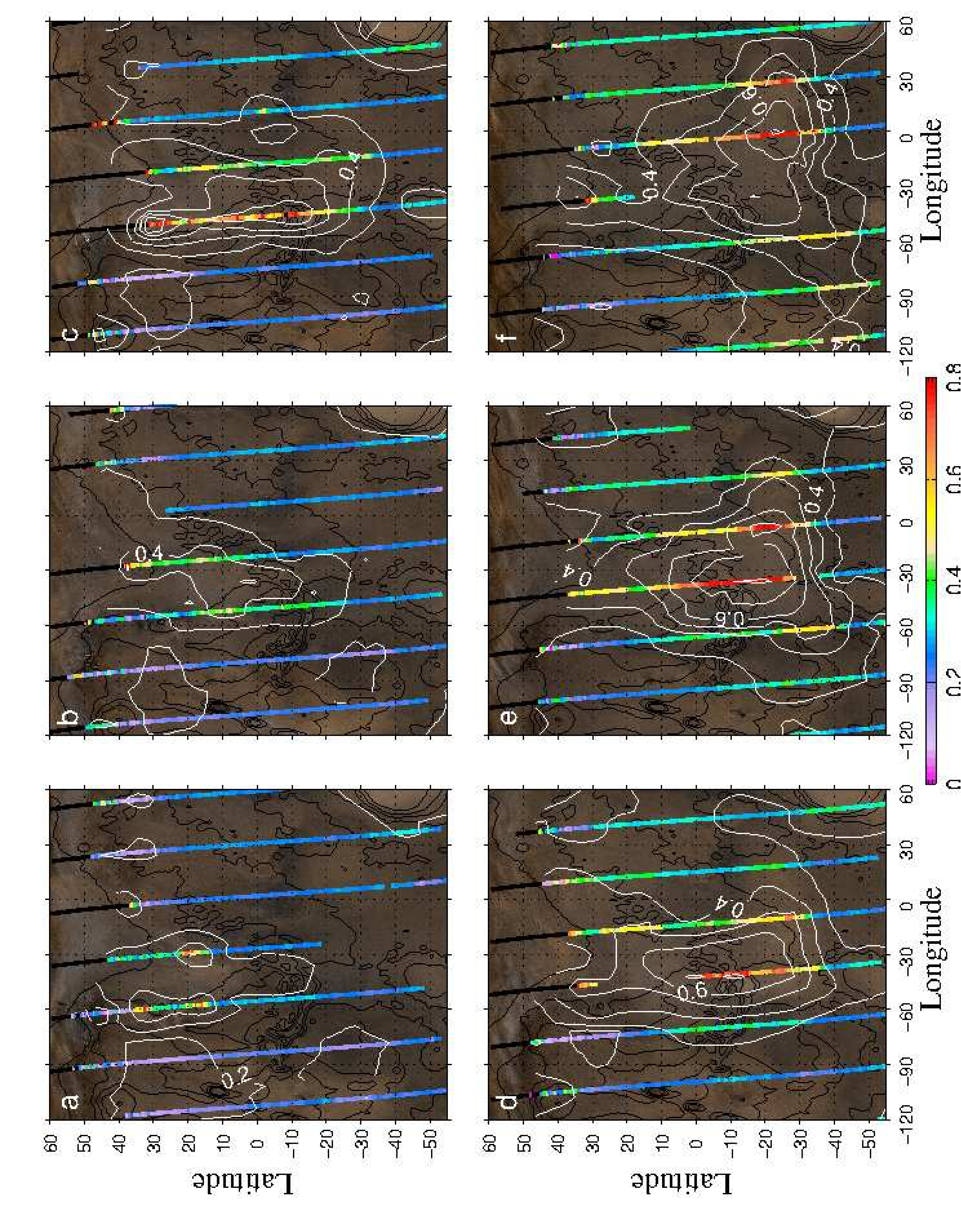}
  \caption{}\label{MOC_TES_MY24}
\end{figure}

\noindent\textbf{Figure~\ref{MOC_TES_MY24}} \,\,Tracks of TES column dust optical depth (at 9.3~$\mu\mathrm{m}$ in absorption) overlaid on MOC Mars Daily Global Map images for 6 sols at 2 sol intervals in MY~24. Contours show the analyzed optical depth from the gridded maps derived from TES observations as described
in the text. (a) Sol 440 ($\mathrm{L_{s}}=221.4^{\circ}$), (b) Sol 442 ($\mathrm{L_{s}}=222.6^{\circ}$), (c)
Sol 444 ($\mathrm{L_{s}}=223.9^{\circ}$),  (d) Sol 446 ($\mathrm{L_{s}}=225.2^{\circ}$),  (e) Sol 448 ($\mathrm{L_{s}}=226.5^{\circ}$),  and (f) Sol 450 ($\mathrm{L_{s}}=227.8^{\circ}$).  In all maps, MUT is noon, and solar longitude values are provided accordingly.  See \ref{calendar} for the
description of the sol-based Martian calendar we use in this paper. MOC images are provided by
Bruce Cantor.

\begin{figure}[H]
\centering
  \noindent\includegraphics[width=\textwidth]{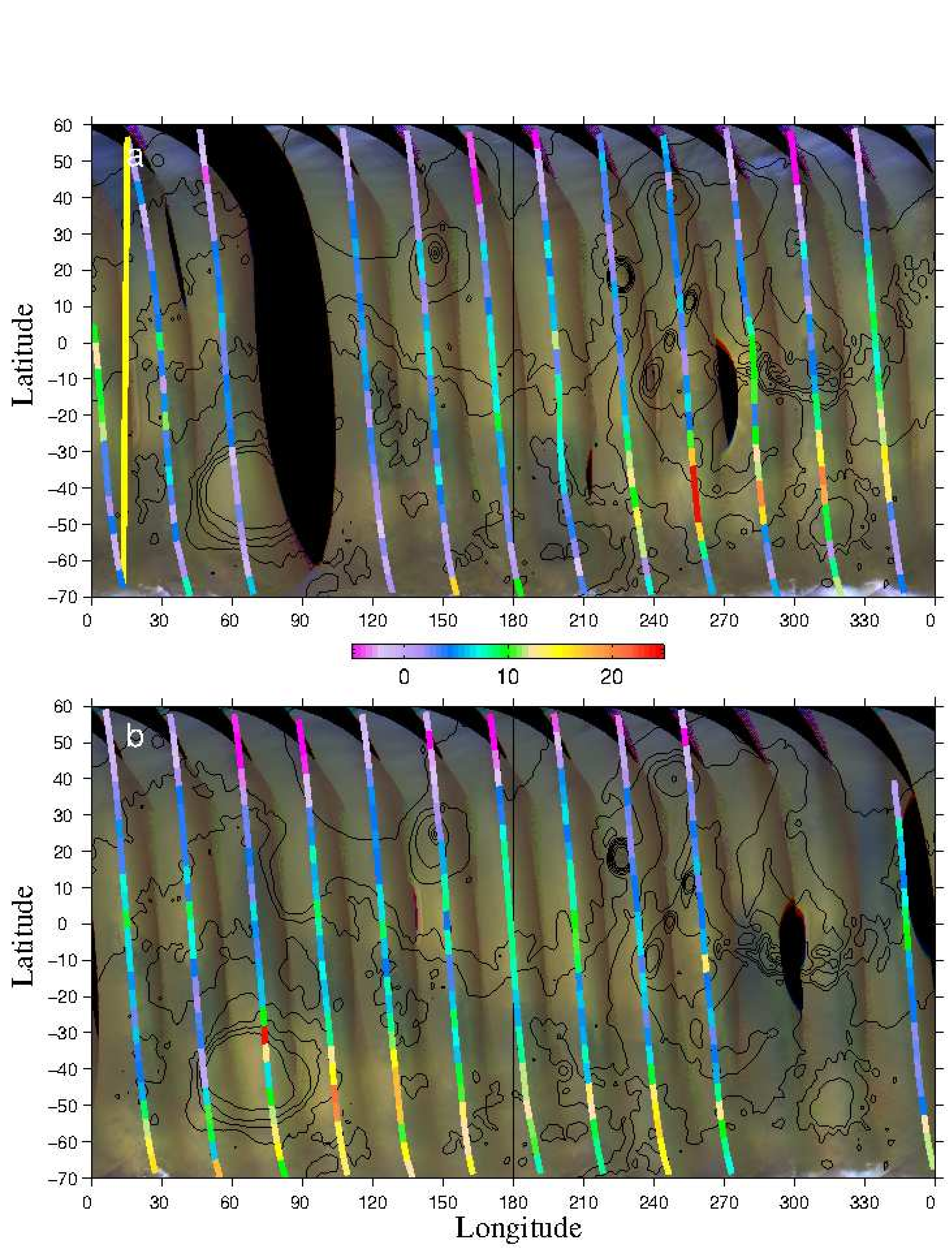}
  \caption{}\label{MARCI_MCS_MY29_3}
\end{figure}

\noindent\textbf{Figure~\ref{MARCI_MCS_MY29_3}} \,\,Tracks of surface temperature anomaly overlaid on MARCI Mars Daily Global Map images for (a) Sol 460 ($\mathrm{L_{s}}=234.2^{\circ}$)  and (b) Sol 468 ($\mathrm{L_{s}}=239.4^{\circ}$). See \ref{calendar} for the description of the sol-based Martian calendar we use in
this paper. Brightness temperature anomaly is the apparent decrease in afternoon
surface temperature (32 micron brightness temperature), relative to a clear
atmosphere, that can be attributed to the influence of dust. MARCI images are provided by Helen Wang.

\begin{figure}[H]
\centering
  \noindent\includegraphics[width=\textwidth]{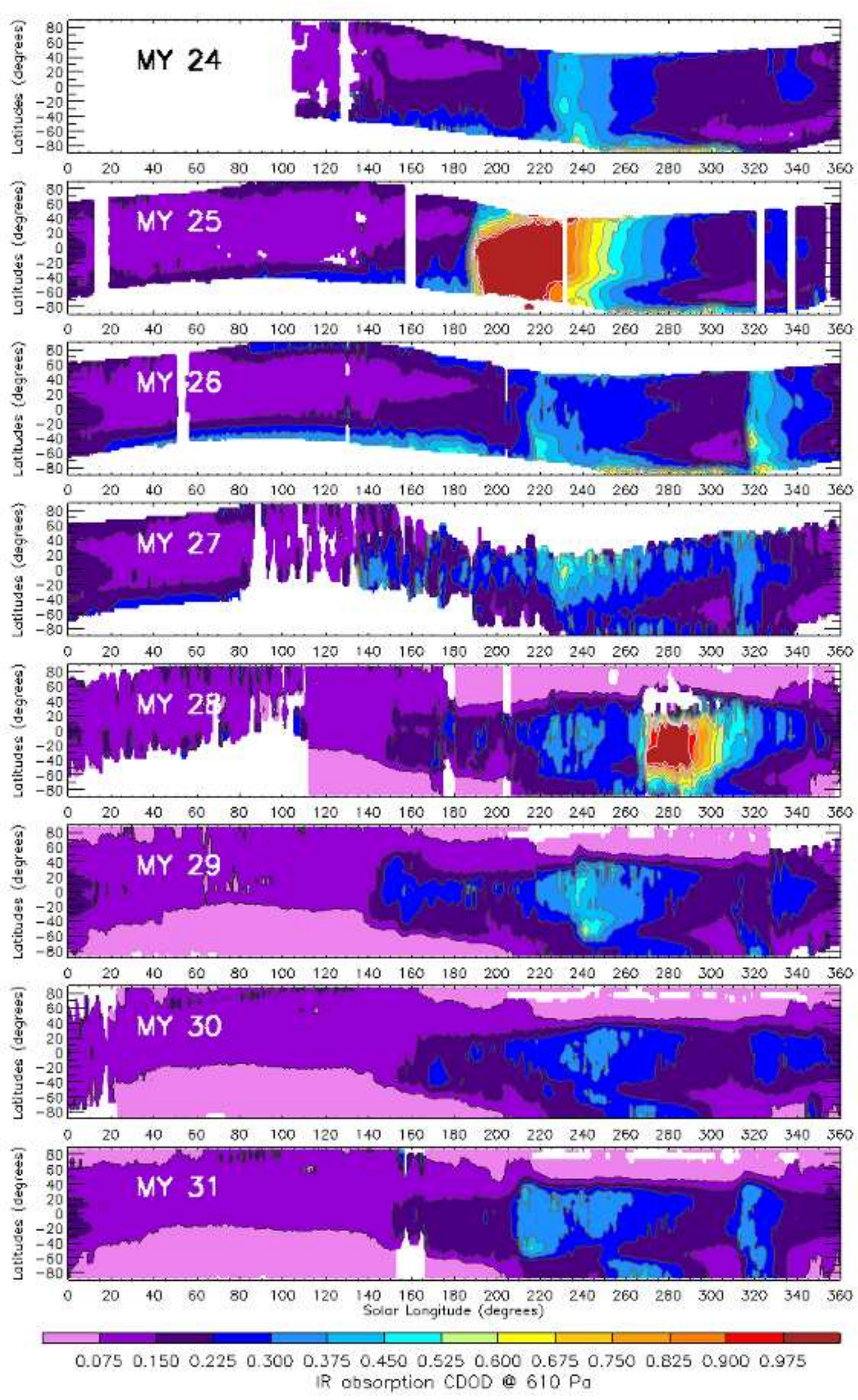}
  \caption{}\label{fig_zonalmean_grid}
\end{figure}

\noindent\textbf{Figure~\ref{fig_zonalmean_grid}} \,\,Zonal means of 9.3~$\mu\mathrm{m}$ absorption column dust optical depth maps (referred to 610 Pa) as a function of solar longitude and latitude for all eight available Martian years. Data are extracted from the irregularly gridded maps obtained with the application of the IWR procedure.

\begin{figure}[H]
\centering
  \noindent\includegraphics[width=\textwidth]{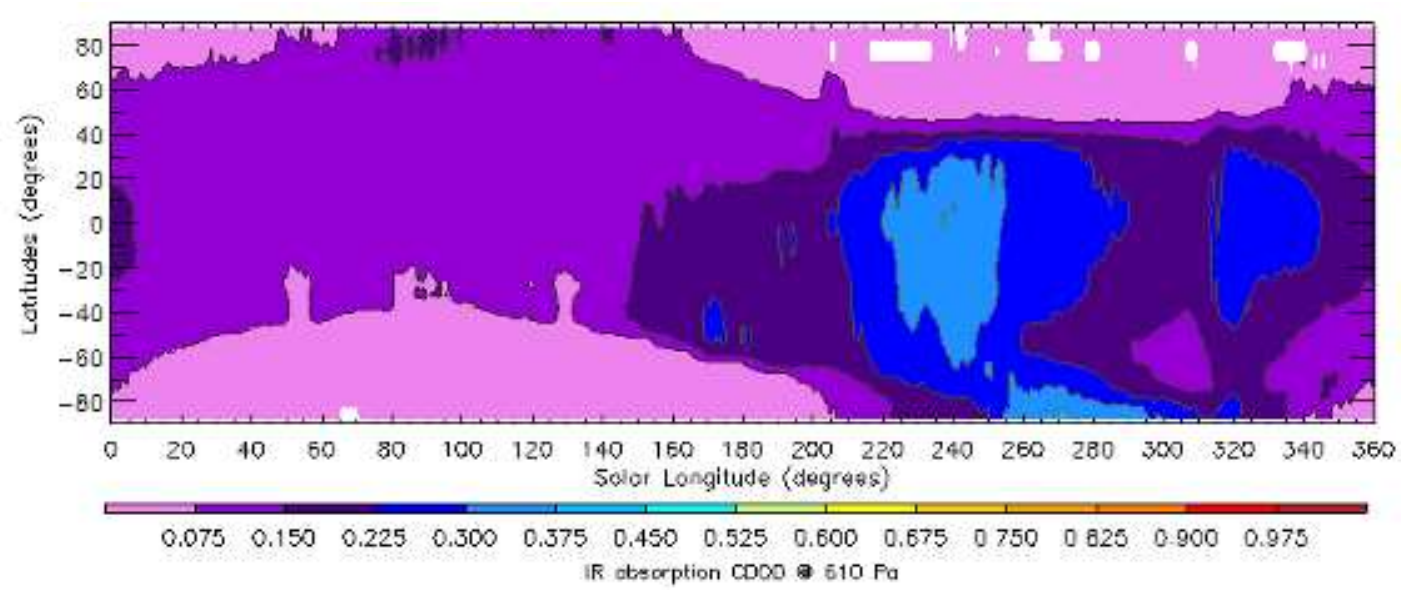}
  \caption{}\label{fig_dodtypyear}
\end{figure}

\noindent\textbf{Figure~\ref{fig_dodtypyear}} \,\,Plot of the zonal mean of the 9.3~$\mu\mathrm{m}$ absorption column dust optical depth in the climatological year, as a function of solar longitude and latitude. The
climatological year is obtained by averaging the irregularly gridded maps for all eight years, excluding the largest value of CDOD for each grid point.

\begin{figure}[H]
\centering
  \noindent\includegraphics[width=\textwidth]{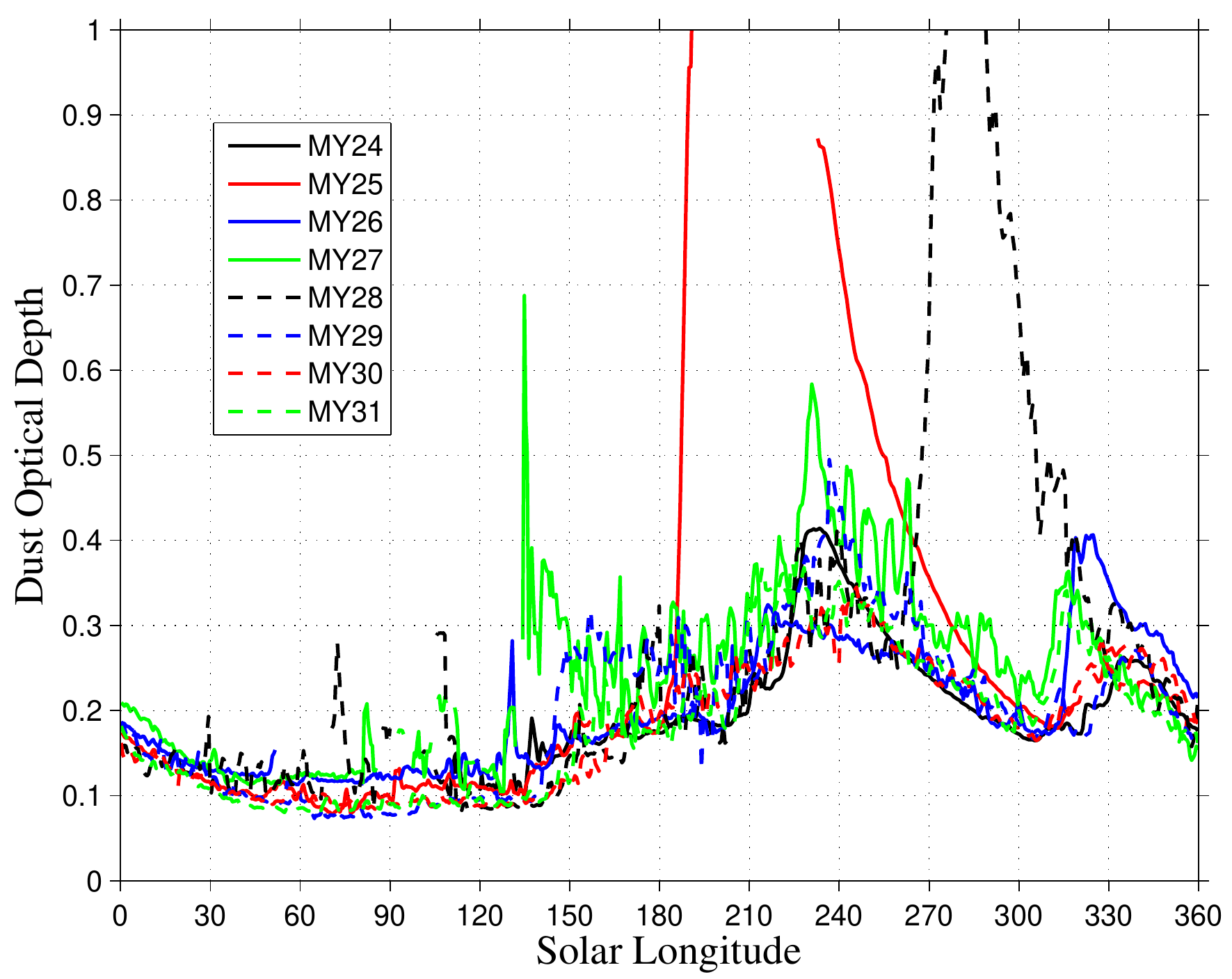}
  \caption{}\label{fig_stat_eq_allyears}
\end{figure}

\noindent\textbf{Figure~\ref{fig_stat_eq_allyears}} \,\,Plot of equatorial ($5^{\circ}$~S--$5^{\circ}$~N) 9.3~$\mu\mathrm{m}$ absorption column dust optical depth as a function of solar longitude for all eight Martian years.

\begin{figure}[H]
\centering
  \noindent\includegraphics[width=\textwidth]{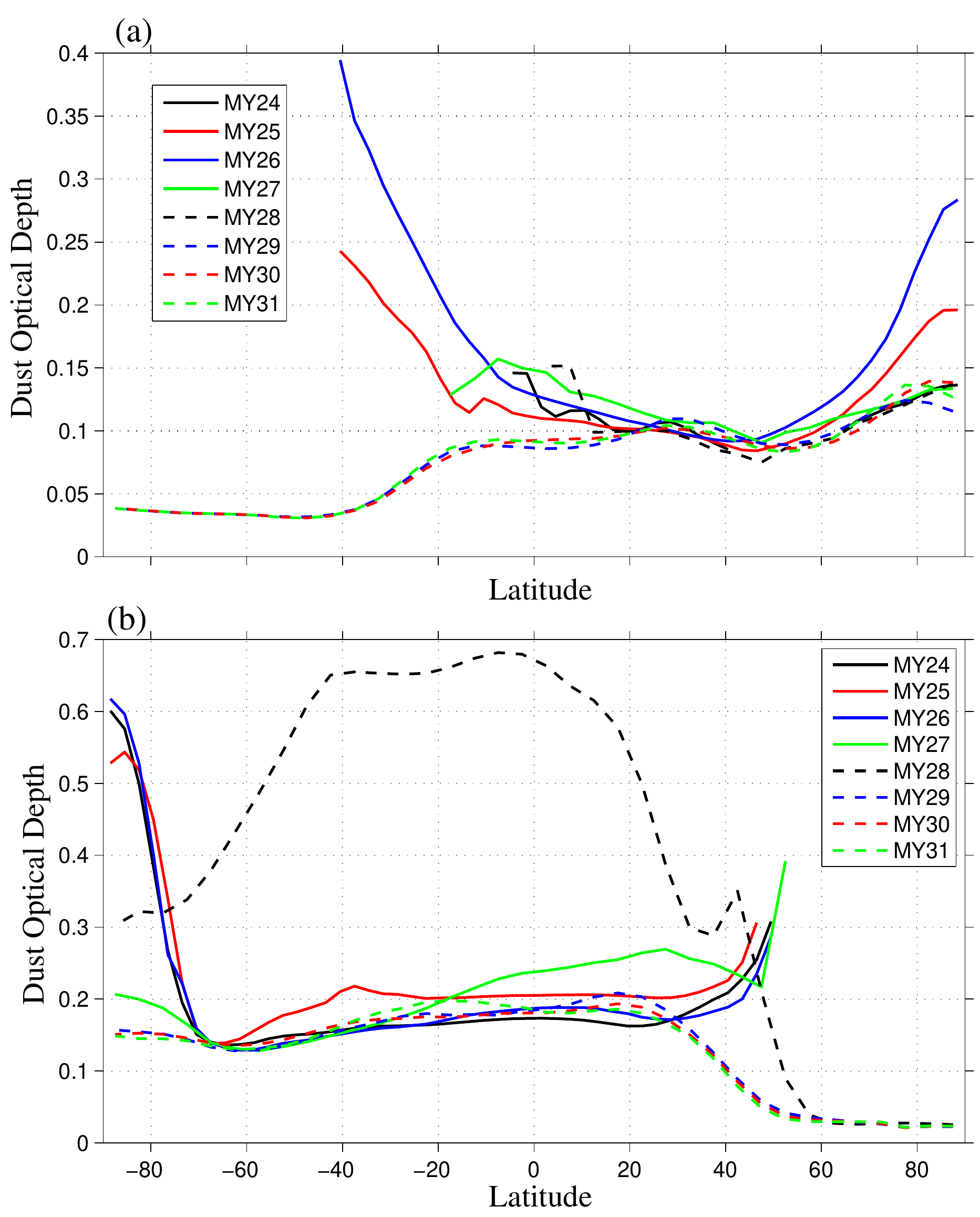}
  \caption{}\label{fig_stat_latallyears}
\end{figure}

\noindent\textbf{Figure~\ref{fig_stat_latallyears}} \,\,Plots of 9.3~$\mu\mathrm{m}$ absorption column dust optical depth as a function of latitude for all eight Martian years, at two seasons: (a) $\mathrm{L_{s}}=[95^{\circ},105^{\circ}]$, and (b) $\mathrm{L_{s}}=[295^{\circ},305^{\circ}]$.

%\noindent\textbf{Figure~\ref{fig_exomars}} \,\,

\begin{figure}[H]
\centering
  \noindent\includegraphics[width=\textwidth]{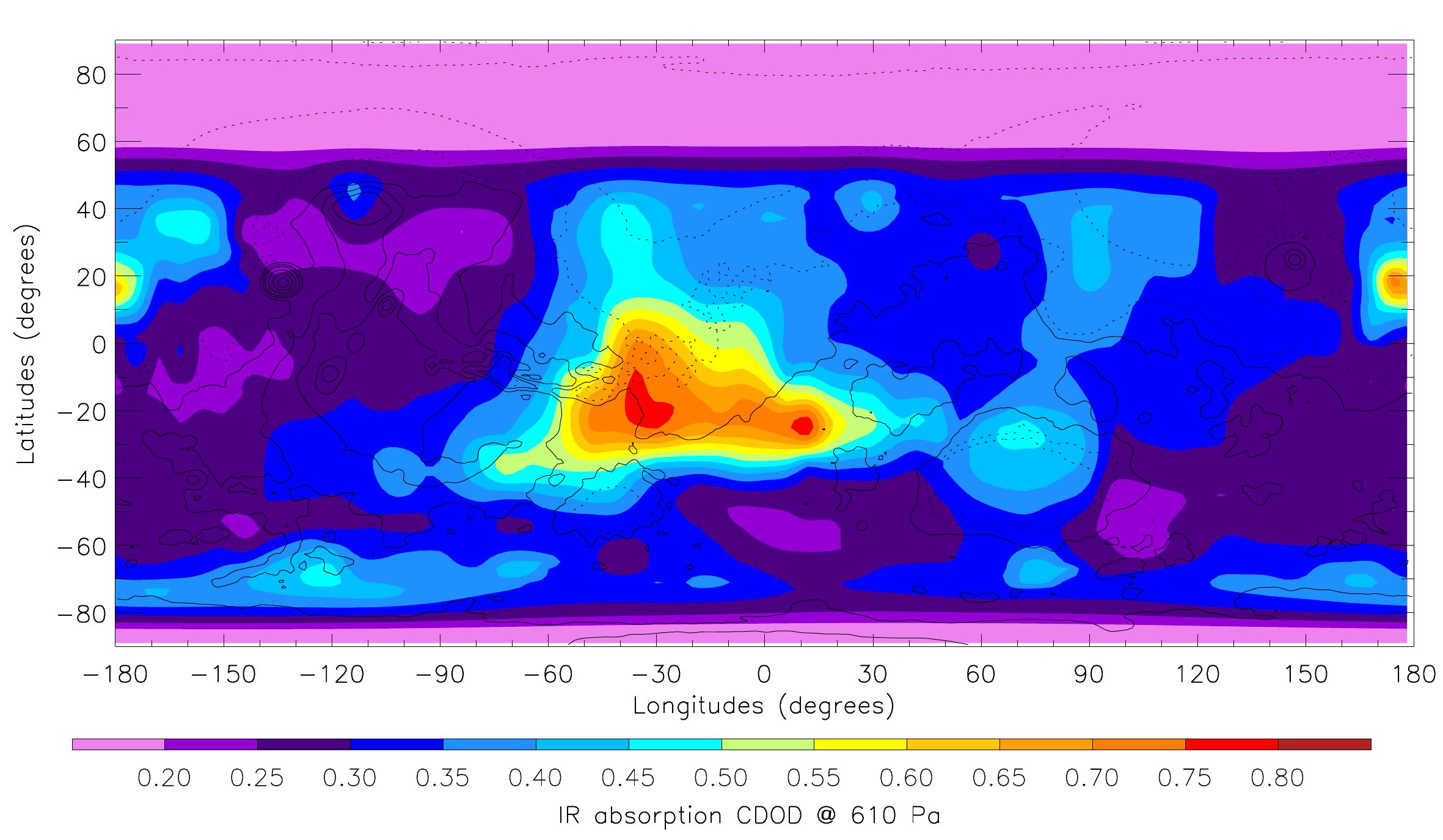}
  \caption{}\label{fig_krigmap}
\end{figure}

\noindent\textbf{Figure~\ref{fig_krigmap}} \,\,Map of 9.3~$\mu\mathrm{m}$ absorption column dust optical depth corresponding to Fig.~\ref{fig_tes_tw} (sol 449, MY~24) after the application of kriging at high
resolution. The map is regular and complete, on a $2^{\circ}\times2^{\circ}$
longitude-latitude grid.

\begin{figure}[H]
\centering
  \noindent\includegraphics[width=\textwidth]{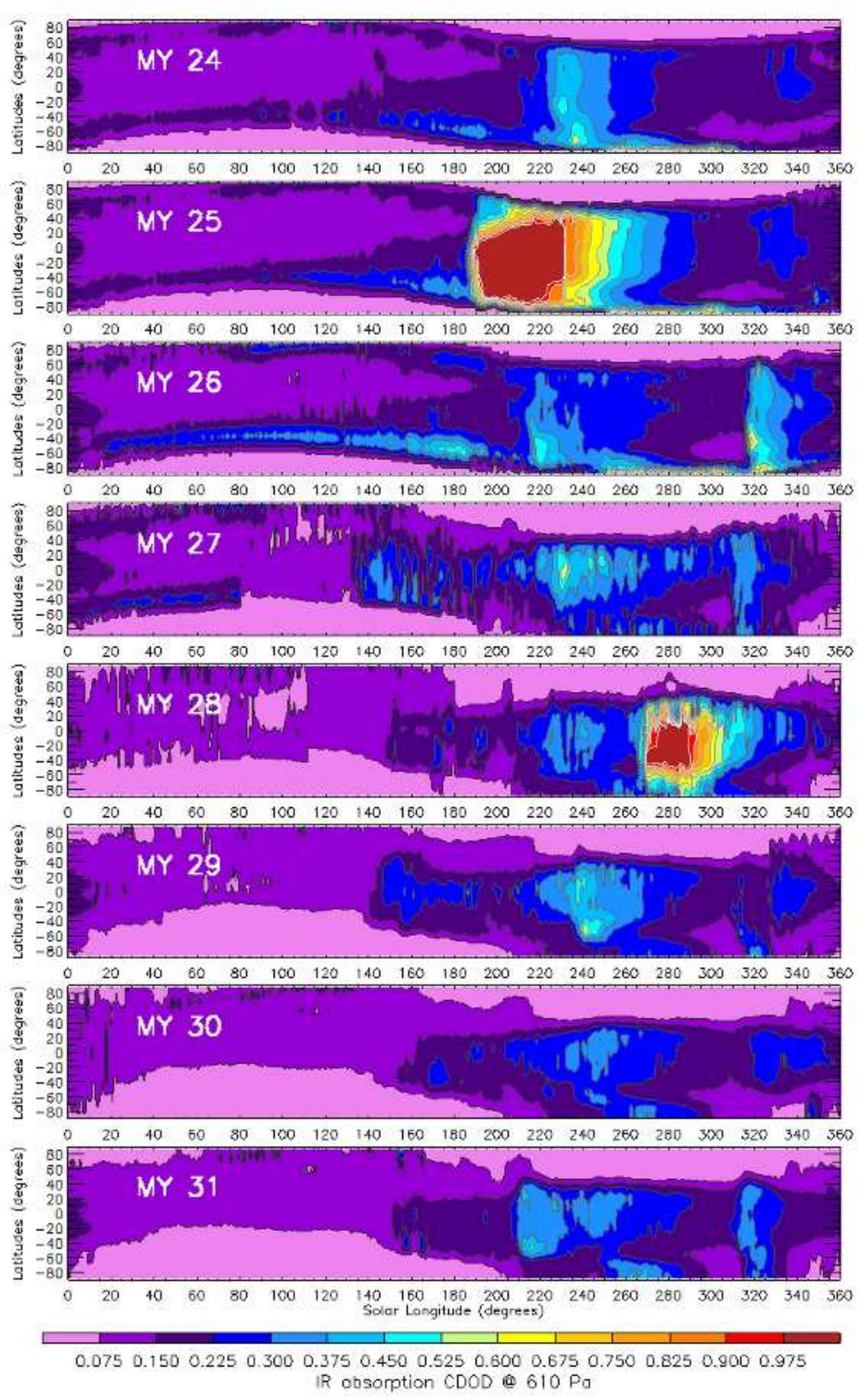}
  \caption{}\label{fig_zonalmean_krig}
\end{figure}

\noindent\textbf{Figure~\ref{fig_zonalmean_krig}} \,\,Zonal means of 9.3~$\mu\mathrm{m}$ absorption column dust optical depth maps (referred to 610 Pa) as a function of solar longitude and latitude for all eight available Martian years,
calculated using the complete maps after the application of the kriging
procedure.

\pagebreak

\pagebreak

\pagebreak

\pagebreak

\pagebreak

\pagebreak

\pagebreak

\pagebreak

\pagebreak

% \pagebreak
% \begin{figure}[H]
% \centering
%   \noindent\includegraphics[width=\textwidth]{fig_correlation_grid_TES}
%   \caption{}\label{fig_correlation_grid_TES}
% \end{figure}
% 
% \pagebreak
% \begin{figure}[H]
% \centering
%   \noindent\includegraphics[width=\textwidth]{fig_correlation_grid_MCS}
%   \caption{}\label{fig_correlation_grid_MCS}
% \end{figure}

\pagebreak

% \pagebreak
% \begin{figure}[H]
% \centering
%   \noindent\includegraphics[width=\textwidth]{fig_correlation_grid_CRISM}
%   \caption{}\label{fig_correlation_grid_CRISM}
% \end{figure}

\pagebreak

\pagebreak

\pagebreak

\pagebreak

\pagebreak

\pagebreak

\pagebreak

\pagebreak

\pagebreak

% \pagebreak
% \begin{figure}[H]
% \centering
%   \noindent\includegraphics[width=\textwidth]{fig_exomars}
%   \caption{}\label{fig_exomars}
% \end{figure}

\pagebreak

\pagebreak

\end{document}